\documentclass[10pt,journal,compsoc]{IEEEtran}
\ifCLASSOPTIONcompsoc
  \usepackage[nocompress]{cite}
\else
\fi
\usepackage[numbers]{natbib}

\usepackage[hidelinks]{hyperref}
\usepackage{amsmath,amssymb,amsfonts}
\usepackage{graphicx}
\usepackage{textcomp}
\usepackage{xcolor}
\usepackage[ruled,vlined]{algorithm2e}
\usepackage{algorithmic}
\usepackage{mathtools}
\usepackage{array}
\usepackage{subcaption}
\usepackage{mwe}
\usepackage{orcidlink}
\usepackage{tikz}
\usepackage{url}
\usepackage{balance}

\usepackage{multirow}
\usepackage{booktabs}
\usepackage{color}
\usepackage{setspace}
\usepackage{lipsum}
\usepackage{verbatim}
\usepackage[inline]{enumitem}

\newtheorem{definition1}{Definition}
\newtheorem{example1}{Example}
\newcommand{\myparagraph}[1]{\vspace{0.3\baselineskip}\noindent{\textbf{#1.}}~}
\ifCLASSINFOpdf
\else
\fi

\begin{document}
%
\title{Dependency Relationships-Enhanced Attentive Group Recommendation in HINs}
%
%
%
%



\author{Juntao Zhang,
        Sheng Wang, Zhiyu Chen,
        Xiandi Yang and Zhiyong Peng

\IEEEcompsocitemizethanks{
\IEEEcompsocthanksitem Juntao Zhang is with the School of Computer and Information Engineering, Henan University, China. Email: juntaozhang@henu.edu.cn.
\IEEEcompsocthanksitem Sheng Wang and Xiandi Yang are
with the School of Computer Science, Wuhan University, China. Email: $\{swangcs, xiandiy\}@$whu.edu.cn.
\IEEEcompsocthanksitem Zhiyu Chen is with Amazon in Seattle, USA. E-mail: zhiyuche@amazon.com.
\IEEEcompsocthanksitem Zhiyong Peng is with the School of Computer Science and Big Data Institute, Wuhan University, China. Email: peng@whu.edu.cn.

}
\thanks{Manuscript received XXX, XXX; revised XXX, XXX.}}

\markboth{IEEE TRANSACTIONS ON KNOWLEDGE AND DATA ENGINEERING,~Vol.~X, No.~X, May~2023}%
{Shell \MakeLowercase{\textit{et al.}}: Bare Demo of IEEEtran.cls for Computer Society Journals}

\IEEEtitleabstractindextext{%
\begin{abstract}
Recommending suitable items to a group of users, commonly referred to as the group recommendation task, is becoming increasingly urgent with the development of group activities.
The challenges within the group recommendation task involve aggregating the individual preferences of group members as the group's preferences and facing serious sparsity problems due to the lack of user/group-item interactions.
To solve these problems, we propose a novel approach called Dependency Relationships-Enhanced Attentive Group Recommendation (\textsc{DREAGR}) for the recommendation task of occasional groups.
Specifically, we introduce the dependency relationship between items as side information to enhance the user/group-item interaction and alleviate the interaction sparsity problem.
Then, we propose a Path-Aware Attention Embedding (\textsc{PAAE}) method to model users' preferences on different types of paths.
Next, we design a gated fusion mechanism to fuse users' preferences into their comprehensive preferences.
Finally, we develop an attention aggregator that aggregates users' preferences as the group's preferences for the group recommendation task.
We conducted experiments on two datasets to demonstrate the superiority of \textsc{DREAGR} by comparing it with state-of-the-art group recommender models.
The experimental results show that \textsc{DREAGR} outperforms other models, especially HR@N and NDCG@N (N=5, 10), where \textsc{DREAGR}
has improved in the range of 3.64\% to 7.01\% and 2.57\% to 3.39\% on both datasets, respectively.

\end{abstract}

}

\maketitle
\IEEEdisplaynontitleabstractindextext
\IEEEpeerreviewmaketitle

\section{Introduction}
\IEEEPARstart{R}{ecommender} systems aim to recommend appropriate items to users via their historical preferences, especially when users are inundated with tremendous information and resources on the internet, which play a significant and indispensable role in alleviating the problem of information overload \cite{ZhangYST19, DaraCK20}.
Recently, group activities have become popular with the development of social networks, which has led to a surge in demand for recommending items to a group of users, namely the group recommendation \cite{Cao0MAYH18, GuoYCZZ22}.
Group recommendation aims to recommend appropriate items that satisfy the demand of a group of users according to the agreement or preferences of group users, which have been applied in various domains such as e-commerce \cite{ZhangGJ021}, tourism \cite{Gross17}, social media \cite{HeCZ20}, and etc. 
There are two types of groups according to the stability of members: persistent groups \cite{Cao0MAYH18, HuCXCGC14} and occasional groups \cite{HeCZ20, SankarWWZYS20}.
Persistent groups have stable members with similar preferences and abundant historical group-item interactions \cite{HuCXCGC14, Cao0MAYH18}, such as a family.
Occasional group (e.g., a travel group) contains a set of ad hoc users (who might join the group for the first time), and its historical group-item interactions are very sparse and even unavailable \cite{QuintarelliRT16, YinWZLZ22}.


In practice, individual recommendation methods can be applied directly to the recommendation task of persistent groups by regarding a group as a virtual user that ignores the distinct preferences of group members \cite{HeLZNHC17, KorenBV09, YuZHG06} based on rich interaction records.
However, this strategy is incompetent for the recommendation tasks of occasional groups.
One reason is the sparse user/group-item interaction, and another reason is that users in a group may have different influences on items.
In this work, we focus on the problem of interaction sparsity and preference aggregation of occasional group recommendations and utilize the dependency relationship between items to enhance user/group-item interactions and model their preferences.

\begin{figure*}[t]
  \includegraphics[width=\linewidth]{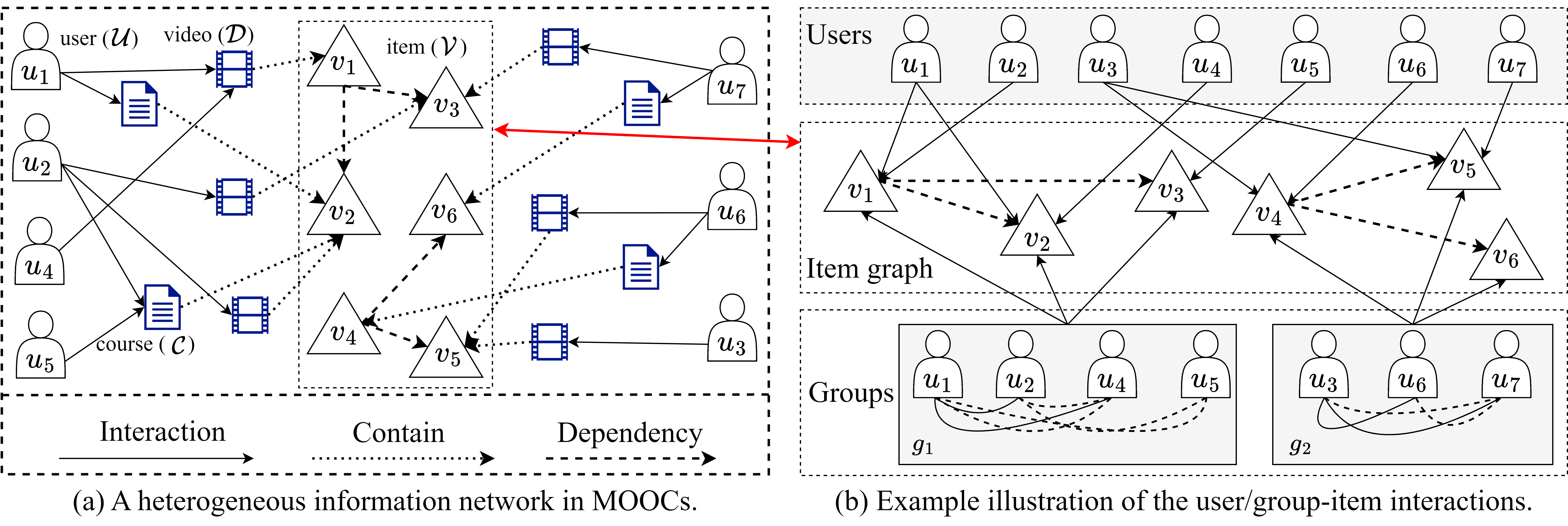}
\vspace{-2.4em}
\caption{A HIN and example of the recommendation task.}
\vspace{-2.0em}
\label{fig:HINE}
\end{figure*}

Several studies have tackled the problem of interaction sparsity by leveraging side information such as social information (e.g., users' social influence) \cite{DelicMNW18, YinW0LYZ19, GuoYW0HC20, YinWZLZ22}, the abundant structure and semantic information of knowledge graphs \cite{DengLLAS21}, etc.
Unlike them, we consider the dependency relationship between items as side information to enhance interaction and alleviate interaction sparsity.
Currently, there are two types of technical solutions to the preference aggregation problem.
One is that memory-based methods aggregate the members' scores (or preferences) using predefined strategies (e.g., average \cite{BaltrunasMR10, BerkovskyF10}, least misery \cite{Amer-YahiaRCDY09}, and maximum satisfaction \cite{BorattoC11}) to represent groups' preferences.
Another is that model-based methods, such as probabilistic models \cite{YeLL12, YuanCL14} and neural attention mechanisms \cite{Cao0MAYH18, TranPTLCL19, SankarWWZYS20, HeCZ20}, model the decision-making process of occasional groups by exploiting the interactions and influences among members.
Unfortunately, probabilistic models assume that users have the same probability in the decision-making with different groups and suffer from high time complexity \cite{GorlaLR013}.
Neural attention mechanisms have been successfully applied in deep learning \cite{Cao0MAYH18, TayLH18}, we consider using it to model users' explicit and implicit preferences and then aggregate them into the group's preferences.

To increase the user/group-item interactions, we model different paths between users and items through heterogeneous information networks (HINs). HINs, consisting of multiple types of entities and their relationships, have been proposed as a powerful information modeling method \cite{ShiHZY19}.
We take the Massive Open Online Courses (MOOCs)~\cite{Gong0WFP0Y20, WangJGL23} as an example, and model entities, such as users, videos, courses, knowledge concepts (here it represents the item), and their relationships as a HIN, as shown in Fig. \ref{fig:HINE}(a).
Unlike the individual recommendations in these two studies, we consider conducting group recommendations in MOOCs and utilizing the dependency relationship between items (e.g., the relationship between $v_1$ and $v_3$) of HINs as side information to alleviate the problem of sparse interaction.
As these do not contain explicit groups in MOOCs, inspired by the strategy for extracting implicit groups in \citet{GuoYCZZ22}, we define meta-paths (Definition \ref{def:mp}) and dependency meta-paths (Definition \ref{def:pmp}) to establish connections between users in HINs and generate implicit groups via following \citet{ZhangW0P23}.
In addition, we model user preferences when interacting with items on meta-paths and dependency meta-paths and aggregate them into group preferences.
In practice, we use the user/group-item interactions on different paths as the input for our group recommendation task, where the item graph is composed of items and their dependency relationships, as shown in Fig. \ref{fig:HINE}(b).

By enhancing interaction and modeling preferences, we propose a Dependency Relationship-Enhanced Attention Group Recommendation (\textsc{DREAGR}) model for the recommendation task of occasional groups.
We innovatively introduce the dependency relationships between items as side information to enhance the implicit interaction between users and items and alleviate the interaction sparsity problem.
In \textsc{DREAGR}, we propose a Path-Aware Attention Embedding (\textsc{PAAE}) method to learn users' preferences for items based on different paths.
Essentially, aggregating user preferences mimics the decision-making process \cite{GuoYW0HC20} that all members of the group reach a consensus to represent the group's preferences.
Then we develop a gated fusion mechanism to fuse users' preferences on different paths as their comprehensive preferences and an attention preference aggregator to aggregate users' overall preferences as groups' preferences.
In short, we introduce the dependency relationship between items to alleviate interaction sparsity and model the preferences of groups on different paths.

The main contributions of our work are summarized
below:
\begin{itemize}
    \item We propose a \textsc{DREAGR} model that utilizes the rich information features of nodes and their relationships in HINs to recommend suitable items to a group of users.
    \item We introduce the dependency relationship between items as side information to enhance the interaction between groups and items and alleviate the problem of sparse interaction.
    \item We propose a Path-Aware Attention Embedding (\textsc{PAAE}) method to learn users' preferences when interacting with items on different types of paths.    
    \item We design a gated fusion mechanism to fuse users' preferences on different types of paths and develop an attention preference aggregator to aggregate users' preferences in the decision-making process of groups.
    \item We conducted experiments on two real datasets and compared \textsc{DREAGR} with several models, and experimental results exhibited the effectiveness of \textsc{DREAGR}.
\end{itemize}

The remainder of this paper is organized as follows.
Section \ref{rw} reviews the related work on group recommendations.
Section \ref{ppd} introduces several preliminaries and defines the problem of group recommendations.
Section \ref{dmodel} introduces the motivation for our group recommendation, and we present the \textsc{DREAGR} model.
The effectiveness of our \textsc{DREAGR} model is evaluated in Section \ref{Exper}. Finally, we conclude our work and introduce future work in Section \ref{conc}.

\section{Related Work}
\label{rw}

Existing studies of group recommendations based on the type of group can be divided into two categories: persistent group \cite{Cao0MAYH18, HuCXCGC14, Wang0RQ0LZ20, TranPTLCL19} and occasional group \cite{HeCZ20, SankarWWZYS20, DengLLAS21, YuanCL14, YinW0LYZ19, YinWZLZ22, GuoYW0HC20, QuintarelliRT16}.
Persistent groups have stable user members (i.e., a family) with similar preferences and abundant historical group-item interactions, occasional groups contain a set of ad hoc users (i.e., a travel group), and group-item interactions are sparse or unavailable \cite{QuintarelliRT16}.
A few studies achieve persistent group recommendation by treating a group as a virtual user \cite{HuCXCGC14, HeLZNHC17, KorenBV09, YuZHG06}, and individual recommendation methods can meet practical needs.
In this section, we introduce the research progress on solving \textbf{interaction sparsity} and \textbf{preference aggregation} in the recommendation task of occasional groups.

\myparagraph{Alleviating interaction sparsity} A few studies introduce side information (such as social information\cite{DelicMNW18, YinW0LYZ19, GuoYW0HC20, YinWZLZ22}, knowledge graphs \cite{DengLLAS21}, etc.) to alleviate the interaction sparsity problem in occasional group recommendations.
For example, \citet{YinW0LYZ19} introduced the notion of personal social influence to quantify the contributions of group members and proposed a deep social influence learning framework based on stacked denoising auto-encoders to exploit and integrate the available user social network information to alleviate the interaction sparsity problem.
Subsequently, \citet{YinWZLZ22} proposed a novel centrality-aware graph convolution module to leverage the social network in terms of homophily and centrality to address the data sparsity issue of user-item interaction.
\citet{DelicMNW18} analyzed the connections between social relationships and social centrality in the group decision-making process and demonstrated that socially central group members are significantly happier with group choice.
\citet{GuoYW0HC20} utilized the relatively abundant user-item and user-user interactions to learn users’ latent features from the items and users they have interacted with, thereby overcoming the sparsity issue of group-item interaction.
\citet{DengLLAS21} used the knowledge graph as the side information to address the interaction sparsity problem and proposed a Knowledge Graph-based Attentive Group recommendation (KGAG) model to learn the knowledge-aware representation of groups.

\vspace{-0.25em}
\myparagraph{{Preference aggregation}}Two types of technical solutions, such as the memory-based and model-based methods \cite{Cao0MAYH18}, have been proposed to solve the problem of preference aggregation in group recommendations.
Memory-based methods employ predefined strategies (e.g., average \cite{BaltrunasMR10, BerkovskyF10}, least misery \cite{Amer-YahiaRCDY09}, and maximum satisfaction \cite{BorattoC11}) to aggregate preferences of group members as the group's preferences in the group decision-making process.
However, since members of occasional groups have different influences and contributions, memory-based methods are inadequate for the complexity and dynamics of the decision-making process.

Distinct from memory-based methods, model-based methods model the preference and influence of members in the group decision-making process using the probabilistic model \cite{YeLL12, YuanCL14, LiuTYL12, RakeshLR16} and neural attention mechanism \cite{Cao0MAYH18, TranPTLCL19, SankarWWZYS20, HeCZ20, DengLLAS21, ChenYLNWW22, WuX0J0ZY23}.
Unfortunately, probabilistic models assume that members of groups have the same influence in the decision-making with different groups and suffer from high time complexity \cite{GorlaLR013}.
More sophisticatedly, the attention mechanism models the group's representation via learning the implicit embeddings of members for group recommendation.
For example, GAME \cite{HeCZ20} modeled the embeddings of users, items, and groups from multiple views using the interaction graph and aggregated the members’ representations as the group representation via the attention mechanism.
MoSAN \cite{TranPTLCL19} dynamically learned different impact weights of users in different groups and considered the interactions between users in the group for group recommendations.
\citet{SankarWWZYS20} proposed a recommender architecture-agnostic framework called Group Information Maximization, which can integrate arbitrary neural preference encoders and aggregators for occasional group recommendation.
\citet{ChenYLNWW22} proposed CubeRec for group recommendation, which adaptively learns group hypercubes from user embeddings with minimal information loss during preference aggregation and measures the affinity between group hypercubes and item points.
In addition, \citet{GuoYW0HC20} modeled the group decision-making process as multiple voting processes and simulated the voting scheme of group members to achieve group consensus via a social self-attention network.
ConsRec \cite{WuX0J0ZY23} revealed the consensus behind groups' decision-making using member-level aggregation, item-level tastes, and group-level inherent interests.

\vspace{0.3em}
\noindent\textbf{\underline{\textit{Remarks}}.}
We consider the following ideas to address the problem of sparse interaction and preference aggregation in the group recommendation task. (1) We introduce natural dependency relationships between items as side information to enhance the user/group-item interaction and alleviate the problem of sparse interaction. (2) We propose a Path-Aware Attention Embedding (PAAE) method to learn users' preferences when interacting with items on different types of paths in HINs and aggregate user preferences into group preferences via a preference aggregator.

\vspace{-0.6em}
\section{Preliminaries and Problem Definition}
\label{ppd}
To represent entities and their relations, such as MOOCs, we model them as heterogeneous information networks (HINs), as shown in Fig. \ref{fig:HINE} (a).
Before defining the recommendation task of occasional groups, we introduce several preliminaries of HINs.
To conveniently understand the symbol of definitions in this work, Table \ref{table:symbol1} lists relevant symbols and their descriptions.

\vspace{-0.8em}
\subsection{Preliminaries}
\begin{definition1}(\textbf{Heterogeneous Information Networks}) \cite{SunHYYW11, Gong0WFP0Y20}
A HIN $\mathcal{H}$ = $(\mathcal{N}, \mathcal{E})$, $\mathcal{N} =  {\textstyle \bigcup_{i=1}^{\left| \mathcal{T} \right|}} N_i$ is a set of nodes, $\mathcal{T}$ = $\{T_1, ..., T_{\left| \mathcal{T} \right|}\}$ is a set of $\left| \mathcal{T} \right|$ entity types in $\mathcal{N}$, and $N_i$ is the node set about the entity type $T_i$.
$\mathcal{E} =  {\textstyle \bigcup_{j=1}^{\left| \mathcal{R} \right|}} E_j$ is a set of edges, $\mathcal{R}$ = $\{R_1, ..., R_{\left| \mathcal{R} \right|}\}$ is a set of $\left| \mathcal{R} \right|$ relation types between entities in $\mathcal{T}$, and $E_j$ is the edge set about the relation types $R_j$. HINs require that $\left| \mathcal{T} \right|$ + $\left| \mathcal{R} \right| >$ 2.
\label{def:hin}
\end{definition1}

\begin{table}
\caption{Descriptions of symbols}
\vspace{-1.2em}
\label{table:symbol1}
\begin{center}
  \begin{tabular}{p{1.6cm} p{6.4cm}}
  \hline
  \textbf{Symbols}&\textbf{Description} \\
  \hline
    $\mathcal{H}$ & The heterogeneous information network\\
    
    $\mathcal{N}, \mathcal{E}$ & The set of nodes $\emph{N}$ and the set of edges $\emph{E}$ \\

    $\phi$, $\psi$ & The entity and relation mapping function \\

    $\mathcal{T}$, $\mathcal{R}$ & The set of node types and relation types \\

    $\vert \mathcal{T} \vert $, $\vert \mathcal{R} \vert$ & The number of node and relation types \\

    $\mathcal{P}$ & The meta-paths \\

    $\mathcal{PP}$ & The dependency meta-paths \\
  \hline
    $\mathcal{U}$, $\mathcal{V}$, $\mathcal{G}$ & The sets of users, items, and groups \\

    $u$, $v$, $g$ & User $u$, item $v$, and group $g$ \\
    $\textbf{Y}^{\mathcal{UV}}$ & The interaction between users and items\\
    $\textbf{Y}^{\mathcal{GV}}$ & The interaction between groups and items \\
    $\textbf{Y}^{\mathcal{VV}}$ & The dependency relationship among items \\
    $\bar{\textbf{p}}_u \in \mathbb{R}^{m}$ & The inherent vector of user $u$ \\
    $\bar{\textbf{q}}_v \in \mathbb{R}^{n}$ & The inherent vector of item $v$ \\
  \hline
  \end{tabular}
\end{center}
\vspace{-2.0em}
\end{table}

\begin{definition1}(\textbf{Network Schema}) \cite{SunHYYW11, Gong0WFP0Y20}
\label{def:ns}
The network schema is a meta template of HINs $\mathcal{H}$ = $(\mathcal{N}, \mathcal{E})$, called $\mathcal{H}_s$ = $(\mathcal{T}, \mathcal{R})$, which shows the relations between entity types through the relation types in $\mathcal{R}$.
The $\mathcal{H}_s$ has two mapping functions: (1) an entity-type mapping function $\phi$: $N \to \mathcal{T}$ maps an entity of $N$ into its types in $\mathcal{T}$; (2) a relation-type mapping function $\psi$: $E \to \mathcal{R}$ maps an edge in $E$ into its relation types in $\mathcal{R}$.
\end{definition1}

\vspace{-0.4em}
Based on the HIN $\mathcal{H}$, we discover two connection paths between users: \textbf{Meta-Paths} and \textbf{Dependency Meta-Paths}.

\begin{definition1}(\textbf{Meta-Paths}) \cite{SunHYYW11}
\label{def:mp}
A meta-path represents a path that connects two entities of the same type via other entity types on the network schema $\mathcal{H}_s$ = $(\mathcal{T}, \mathcal{R})$, denoted as $\mathcal{P}: T_1 \stackrel{R_1}{\longrightarrow} T_2 \stackrel{R_2}{\longrightarrow} ... \stackrel{R_l}{\longrightarrow} T_{l+1}$.
Meta-path $\mathcal{P}$ describes a composite relation $\textsf{R}$ = $R_1 \circ R_2 \circ ... \circ R_l$ between $T_1$ and $T_{l+1}$, where $\circ$ is the composition operator on relations, and $T_1$ and $T_{l+1}$ are the same entity types.
\end{definition1}

\begin{example1}
\label{example:1}
If users $u_i$ and $u_j$ access the same item $v_k$, they relate via the composite relation $\textsf{R}$ in $\mathcal{H}$, and there is a path between them called a \textbf{path instance} $p_{u_i \leadsto u_j}$ of $\mathcal{P}$.
Thus we acquire all path instances based on the meta-path $\mathcal{P}$, denoted as $p_{u_i \leadsto u_j} \vdash \mathcal{P}$.
In Fig. \ref{fig:HINE}(a), we select three types of meta-paths to model the connection path between users in $\mathcal{H}$ by accessing the items, including $\mathcal{P}_1 \; (\mathcal{U} \stackrel{1}{\rightarrow} \mathcal{V} \stackrel{1}{\leftarrow} \mathcal{U})$, $\mathcal{P}_2 \; (\mathcal{U} \stackrel{1}{\rightarrow} \mathcal{D} \stackrel{1}{\rightarrow} \mathcal{V} \stackrel{1}{\leftarrow} \mathcal{D} \stackrel{1}{\leftarrow} \mathcal{U})$,
and $\mathcal{P}_3 \; (\mathcal{U} \stackrel{1}{\rightarrow} \mathcal{C} \stackrel{1}{\rightarrow} \mathcal{V} \stackrel{1}{\leftarrow} \mathcal{C} \stackrel{1}{\leftarrow} \mathcal{U})$, where 1 denotes there is a relationship between entities, the $\mathcal{D}$ and $\mathcal{C}$ are the sets of videos and courses of HINs in Fig. \ref{fig:HINE}(a), respectively.
For example, the users $u_1$ and $u_2$ can establish an association via the item $v_1$ in Fig. \ref{fig:HINE}(b).
At the same time, users can also establish interaction with the central items on meta-paths, such as user $u_1$ (or $u_2$) can interact with item $v_1$.
Therefore, users can establish one-hop interaction with the item based on $\mathcal{P}_{1}$ and two-hop interaction with items based on $\mathcal{P}_{2}$ or $\mathcal{P}_{3}$.
\end{example1}

\begin{definition1}(\textbf{Dependency Meta-Paths}) \cite{ZhangW0P23}
\label{def:pmp}
A dependency meta-path denotes a path that connects two entities of the same type via the dependency relationship of other entity types on the network schema $\mathcal{H}_s$ = $(\mathcal{T}, \mathcal{R})$, denoted as $\mathcal{PP}$:
$T_1 \stackrel{R_1}{\longrightarrow} ... T_i \stackrel{PR_i}{\longmapsto} T_j ... \stackrel{R_l}{\longrightarrow} T_{l+1}$.
$\mathcal{PP}$ connects the same type of entities $T_1$ and $T_{l+1}$ based on a composite relation $\textsf{R}^{p}$ = $R_1^{p} \circ ... \circ R_i^{p} \circ ... \circ R_l^{p}$, where $\circ$ is the composition operator on relations.
The $R_i^{p}$ denotes the dependency relationship between $T_i$ and $T_j$.
\end{definition1}

\begin{example1}
\label{example:2}
Suppose that users $u_i$ and $u_j$ access the items $\emph{v}_{i'}$ and $\emph{v}_{j'}$, respectively, and there is a dependency relationship (e.g., prerequisite relationship) between $\emph{v}_{i'}$ and $\emph{v}_{j'}$.
The users $u_i$ and $u_j$ are related by the composite relation $\textsf{R}_p$ in $\mathcal{H}$, there is a dependency path between them, and we say it is a \textbf{dependency path instance} $p_{u_i \leadsto ... \emph{v}_{i'} \mapsto \emph{v}_{j'} ... \leadsto u_j}$ of $\mathcal{PP}$.
We obtain all dependency path instances based on dependency meta-paths $\mathcal{PP}$, denoted as $p_{v_i \leadsto ... v_{i'} \mapsto v_{j'} ... \leadsto v_j} \vdash \mathcal{PP}$.
In Fig. \ref{fig:HINE}(a), we select three types of dependency meta-paths to denote the connections among users in $\mathcal{H}$ by accessing the items, including $\mathcal{PP}_{1} \; (\mathcal{U} \stackrel{1}{\rightarrow} \mathcal{V}_{i'} \mapsto \mathcal{V}_{j'} \stackrel{1}{\leftarrow} \mathcal{U})$,
$\mathcal{PP}_{2} \; (\mathcal{U} \stackrel{1}{\rightarrow} \mathcal{D} \stackrel{1}{\rightarrow} \mathcal{V}_{i'} \mapsto \mathcal{V}_{j'} \stackrel{1}{\leftarrow} \mathcal{D} \stackrel{1}{\leftarrow} \mathcal{U})$,
and $\mathcal{PP}_{3} \; (\mathcal{U} \stackrel{1}{\rightarrow} \mathcal{C} \stackrel{1}{\rightarrow} \mathcal{V}_{i'} \mapsto \mathcal{V}_{j'} \stackrel{1}{\leftarrow} \mathcal{C} \stackrel{1}{\leftarrow} \mathcal{U})$,
where $\mapsto$ represents the dependency relationship between $\mathcal{V}_{i'}$ and $\mathcal{V}_{j'}$.
For example, the users $u_2$ and $u_4$ can establish an association via the items $v_1$, $v_2$, and their dependency relationship in Fig. \ref{fig:HINE}(b).
Therefore, users can establish two-hop interaction with the item based on $\mathcal{PP}_{1}$ and three-hop interaction with items based on $\mathcal{PP}_{2}$ or $\mathcal{PP}_{3}$.
For example, user $u_1$ establishes two-hop interaction with $v_2$ by the dependency relationship between $v_1$ and $v_2$.
We uniformly represent a two-hop or three-hop interaction between users and items on dependency meta-paths as the multi-hop interaction.
\end{example1}
\vspace{-1.5em}

\subsection{Problem Definition}
Following previous work \cite{Cao0MAYH18, GuoYCZZ22, SankarWWZYS20}, we also use bold lowercase letters (e.g., $\textbf{x}$) and bold capital letters (e.g., $\textbf{X}$) to represent vectors and matrices, respectively.
We utilize non-bold lowercase letters (e.g., $\emph{x}$) and bold capital letters (e.g., $\emph{X}$) to denote scalars.
Note that all vectors are in column forms if not clarified.
The problem definition for the group recommendation task is as follows:

\textbf{Input:} Users $\mathcal{U}$, items $\mathcal{V}$, groups $\mathcal{G}$, the user-item interactions $\textbf{Y}^{\mathcal{UV}}$, the group-item interactions $\textbf{Y}^{\mathcal{GV}}$, and the item-item dependency relationship $\textbf{Y}^{\mathcal{VV}}$.

\textbf{Output:} 
A function $\mathcal{F}$ that maps the probability between a group and an item: $\hat{y}_{gv}=\mathcal{F}(g,v|\Theta, \textbf{Y}^{\mathcal{UV}}, \textbf{Y}^{\mathcal{GV}}, \textbf{Y}^{\mathcal{VV}})$. We determine whether to recommend the item to the group based on $\hat{y}_{gv}$, where $\Theta$ is the parameter set of the $\mathcal{F}$.

Fig. \ref{fig:HINE}(b) shows the input of the group recommendation task that our work will finish.
Let $\mathcal{U} = \{u_1,...,u_n\}$, $\mathcal{V} = \{v_1,...,v_m\}$, and $\mathcal{G} = \{g_1,...,g_s\}$ be the sets of users, items, and groups, where $n$, $m$, and $s$ are the numbers of elements in these three sets, respectively.
The $l$-th group $g_l \in \mathcal{G}$ consists of a set of users, such as $\mathcal{G}(l) = \{u_1^{l}, u_2^{l},...,u_{|g_l|}^{l} \}$, where $\left | g_l \right |$ is the size of $g_l$ and $\mathcal{G}(l)$ denotes the user set of $g_l$.
We denote each user $u \in \mathcal{U}$ and item $v \in \mathcal{V}$ inherent vectors in HINs using $\bar{\textbf{p}}_u \in \mathbb{R}^{m}$ and $\bar{\textbf{q}}_v \in \mathbb{R}^{n}$, respectively.

There are three kinds of intuitive relationships among $\mathcal{U}$, $\mathcal{V}$, and $\mathcal{G}$ in Fig. \ref{fig:HINE}(b), called the user-item interaction relationship (denoted as $\textbf{Y}^{\mathcal{UV}} = \left [\textbf{y}_{i,j}^{uv}  \right ]^{n \times m}$), the group-item interaction relationship ($\textbf{Y}^{\mathcal{GV}} = \left [\textbf{y}_{k,j}^{gv}  \right ]^{s \times m}$), and the item-item dependency relationship ($\textbf{Y}^{\mathcal{VV}} = \left [\textbf{y}_{j,j}^{vv}  \right ]^{m \times m}$).
We obtain $\textbf{y}^{gv}$ through the user $u$ of a group $g$ and the interaction between the user $u$ and the item $v$. Note that $y=1$ indicates the existence of an interaction relationship, and $y=0$ means there is no interaction relationship.
The item-item relationship indicates an inherent dependency relationship between items, such as the prerequisite relationship among knowledge concepts in the education domain.
Additionally, we can obtain implicit multi-hop interaction $\textbf{Y}^{\mathcal{UVV}}$ between users and items through $\textbf{Y}^{\mathcal{UV}}$ and $\textbf{Y}^{\mathcal{VV}}$.
Similar to $\textbf{Y}^{\mathcal{UVV}}$, we can obtain implicit multi-hop interaction $\textbf{Y}^{\mathcal{GVV}}$ between groups and items.


\section{DREAGR Model}
\label{dmodel}
\begin{figure*}
\centering
  \includegraphics[width=\linewidth]{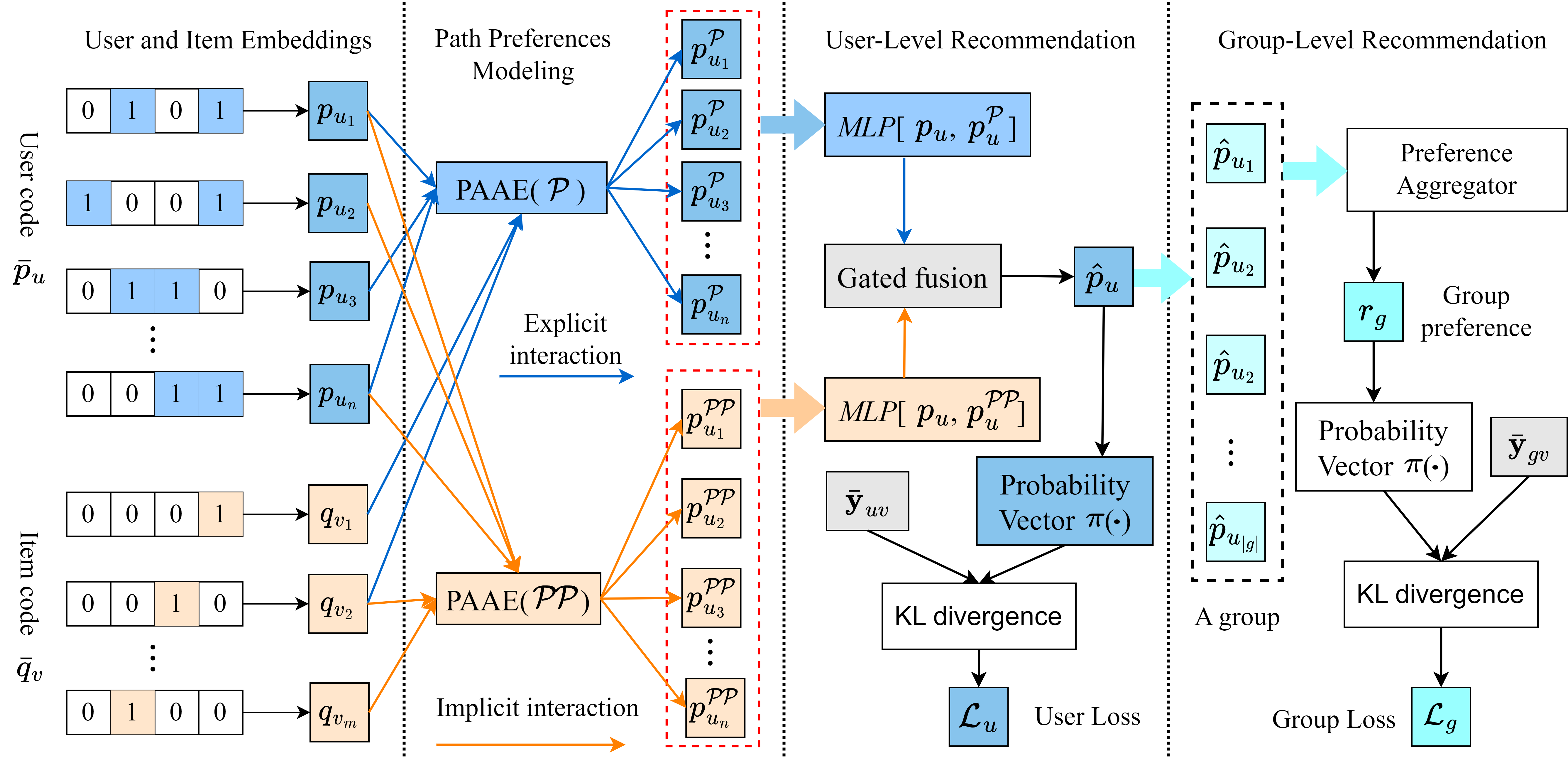}
\vspace{-2.0em}
\caption{The architecture of \textsc{DREAGR}.}
\label{fig:model}
\vspace{-1.8em}
\end{figure*}

In this section, we introduce our \textsc{DREAGR} model for the recommendation task of occasional groups.
The architecture of our \textsc{DREAGR} model is shown in Fig. \ref{fig:model}, which contains four key components.
We first introduce the motivation of our \textsc{DREAGR} model to address the group recommendation task.
Then, we will present a detailed introduction to the content of each component in our \textsc{DREAGR} model and its training optimization.

\subsection{Motivation}
\begin{figure}
\centering
  \subfloat[\textbf{MOOCCube}]{
    \centering
	\includegraphics[scale=0.156]{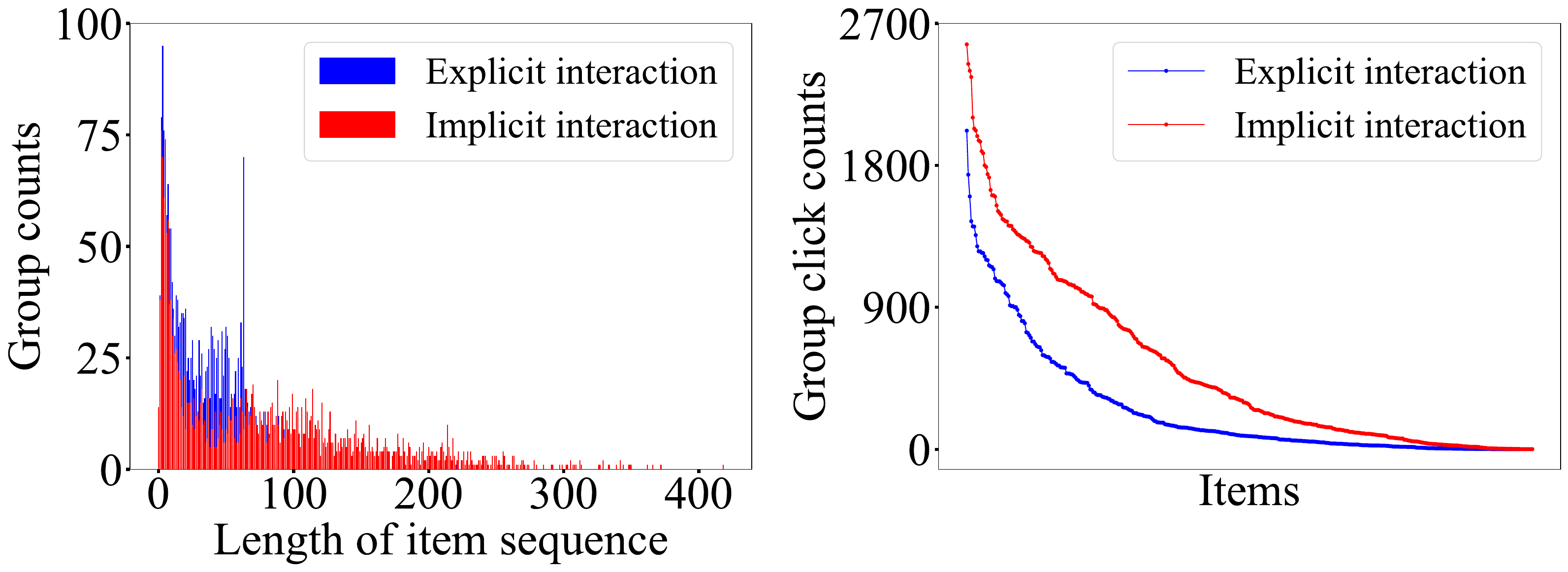}}
	\\
  \subfloat[\textbf{Movielens}]{
      \centering
	  \includegraphics[scale=0.156]{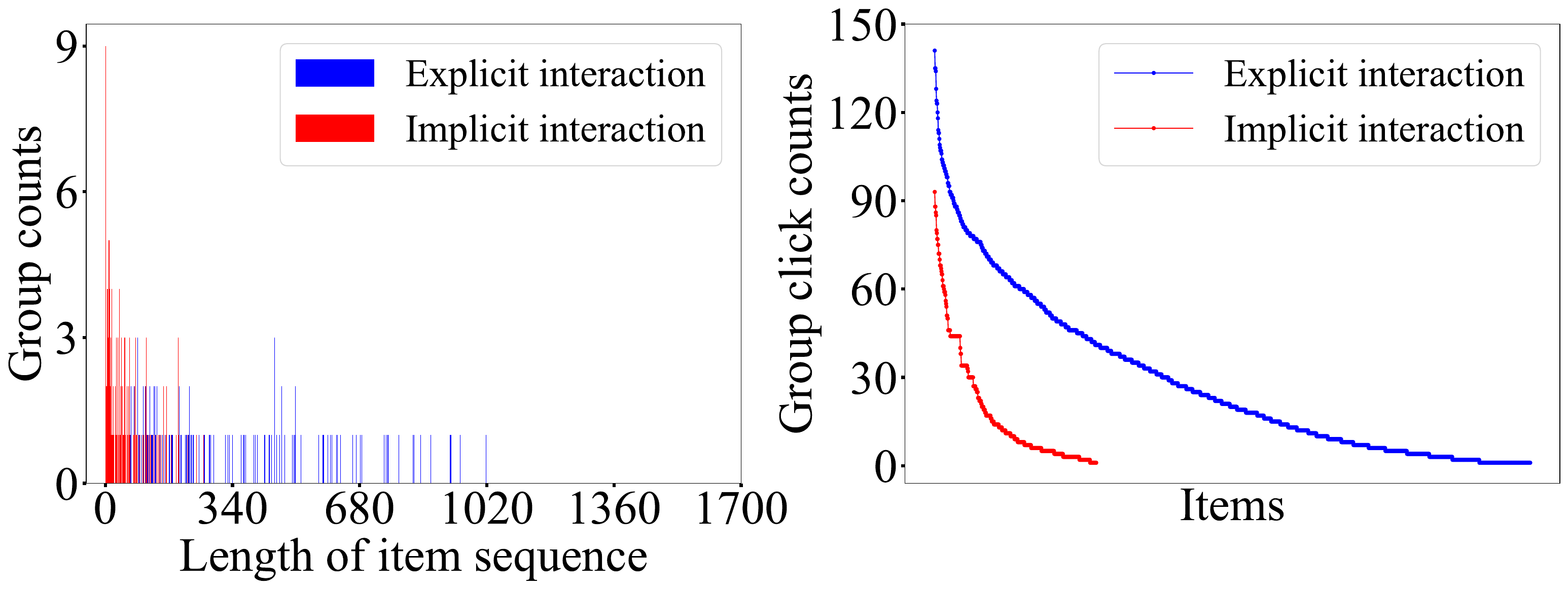}}
\vspace{-1.0em}
\caption{The statistical information of group-item interactions.}
\label{fig:stat}
\vspace{-1.8em}
\end{figure}

In practice, the problem of sparse interaction is a critical problem that the recommendation task of occasional groups needs to address.
An intuitive way to solve this problem is to increase user/group-item interaction.
Naturally, users or groups can establish connections with items on meta-paths and dependency meta-paths in HINs.
We calculate the information of group-item interactions according to the user-item interactions on meta-paths and dependency meta-paths, denoted as \emph{Explicit interaction} and \emph{Implicit interaction}, respectively.
We present the group-item interaction information on our two datasets, as shown in Fig. \ref{fig:stat}.

Both the \emph{Explicit} and \emph{Implicit} interactions show the ``Long-tail Effect'', and more than half of the items have little interaction with groups.
Compared to \emph{Explicit interaction} on the \textbf{MOOCCube} dataset, \emph{Implicit interaction} establishes the multi-hop interaction (the two-hop or three-hop interactions on different paths) between groups and items using the dependency relationship between items, which increases the number of interactions between groups and item sequences and significantly alleviates the ``Long-tail Effect''.
However, \emph{Implicit interaction} is inferior to \emph{Explicit interaction} because there are fewer dependency meta-paths in the \textbf{Movielens} dataset.
We propose integrating \emph{Explicit} and \emph{Implicit} interactions to increase the number of interactions to alleviate the problem of sparse interactions.
This is one of our motivations via the dependency relationship between items as side information to alleviate the problem of sparse interaction for implementing group recommendation tasks.




Users' preference aggregation in group decision-making is also a critical problem of group recommendations.
In HINs, users can associate with other users through different items on meta-paths (Definition \ref{def:mp}) and dependency meta-paths (Definition \ref{def:pmp}).
Different types of paths usually convey different semantics, which indicates that users have unique preferences on these paths.
Therefore, we propose a Path-Aware Attention Embedding (\textsc{PAAE}) method to learn users' preferences for items on meta-paths and dependency meta-paths, respectively, called the user's explicit and implicit preferences.
Then, we develop a gated fusion mechanism to fuse users' preferences on different types of paths and inherent features of users into their comprehensive preferences.
Finally, we design an attention preference aggregator to aggregate users' comprehensive preferences as groups' preferences in the group decision-making process.
Another one of our motivations for implementing group recommendation tasks is to fuse users' preferences on different types of paths and aggregate them into the group's preferences.

\subsection{User and Item Embeddings}
In the \textsc{DREAGR} model, we encode users and items using \emph{One-Hot Encoding} to represent their inherent features, i.e., their actual interaction.
Thus, we use $\bar{\textbf{p}}_u \in \mathbb{R}^{m}$ and $\bar{\textbf{q}}_v \in \mathbb{R}^{n}$ to denote the initial vectors of the user $u$ and item $v$, respectively, where \emph{n} and \emph{m} are the numbers of users and items.
To improve the efficiency of the \textsc{DREAGR} model in the training process, we convert the high-dimensional vectors $\bar{\textbf{p}}_u$ and $\bar{\textbf{q}}_v$ of the user $u$ and item $v$ into their corresponding low-dimensional embeddings via a fully connected network (\emph{FCN}), denoted as $\textbf{p}_u \in \mathbb{R}^{F}$ and $\textbf{q}_v \in \mathbb{R}^{F}$, respectively.
Their calculations are represented as follows:
\begin{equation}
   \textbf{p}_u = FCN(\textbf{W}_u \bar{\textbf{p}}_u + \textbf{b}_u),
\end{equation}
\begin{equation}
   \textbf{q}_v = FCN(\textbf{W}_v \bar{\textbf{q}}_v + \textbf{b}_v),
\end{equation}
where $\textbf{W}_u \in \mathbb{R}^{F \times m}$ and $\textbf{W}_v \in \mathbb{R}^{F \times n}$ are weight matrices of the user $u$ and item $v$, respectively, $\textbf{b}_u \in \mathbb{R}^{F}$ and $\textbf{b}_v \in \mathbb{R}^{F}$ are their corresponding bias vectors.

\subsection{User Preferences Modeling}
In HINs, a user associates with other users through different items on meta-paths and dependency meta-paths, which indicates that the user-item interactions form users' unique preferences.
In the \textsc{DREAGR} model, we propose a Path-Aware Attention Embedding method (\textsc{PAAE}) to model user preference representations on different types of paths.

\subsubsection{Modeling Users' Explicit Preferences}
In HINs, we can establish the association between users via meta-paths.
Therefore, we can also establish the interaction between users and items to form users' preference representations, as the central entity of the meta-path is the item, called the users' explicit preferences.
For example, user $u_1$ can be associated with user $u_2$ on the meta-path instance $u_1 \stackrel{1}{\rightarrow} v_1 \stackrel{1}{\leftarrow} u_2$, indicating that user $u_1$ (or $u_2$) has some preference for item $v_1$.
In addition, user $u_1$ can be associated with user $u_4$ on the meta-path instance $u_1 \stackrel{1}{\rightarrow} v_2 \stackrel{1}{\leftarrow} u_4$, indicating that user $u_1$ also has some preference for item $v_2$.
In practice, there are some differences in users' preferences because the content and number of items they interact with are different on meta-path instances in HINs.
Consequently, it is valuable for us to derive the users' explicit preference representations from the view of meta-paths via the user-item interaction.

We leverage \textsc{PAAE} to model the user-item interactions of user $u$ on all meta-paths in HINs and denote his (her) preference representation as $\textbf{p}_{u}^{\mathcal{P}}$.
For the preference representation of user $u$ on a type of the meta-path $\mathcal{P}_l$, we denote it by $\textbf{p}_{u}^{\mathcal{P}_l} \in \mathbb{R}^{F}$ and calculate as follows:
\begin{equation}
   \textbf{p}_{u}^{\mathcal{P}_l} = \textsc{PAAE}(\textbf{Y}^{\mathcal{UV}}, \mathcal{P}_l, u) = \sum_{j\in Y_{\mathcal{V}}(u)} \alpha_{j}^{\mathcal{P}_l} \textbf{q}_j,
\label{equ:equ3}
\end{equation}
where the $Y_{\mathcal{V}}(u)$ denotes the set of items that the user $u$ interacts with on the meta-path $\mathcal{P}_l$, the $\textbf{q}_j$ is the embedding of the item $j$ ($v_j$ is abbreviated as $j$).
The $\alpha_{j}^{\mathcal{P}_l}$ is the attention weight of user $u$ to the item $j$ from the perspective of the meta-path $\mathcal{P}_l$, which indicates the importance of different items to the user $u$.
In reality, if a user has a high weight on an item, he (she) has more influence in groups \cite{QuintarelliRT16}.
Thus, we input the embedding $\textbf{p}_u$ of user $u$ and the embedding $\textbf{q}_j$ of item $j$ on the meta-path $\mathcal{P}_l$ into an attention model to calculate the attention weight $\alpha_{j}^{\mathcal{P}_l}$, as follows:
\begin{equation}
   e_{j}^{\mathcal{P}_l} = (\textbf{h}_{\mathcal{P}_l}^{v})^{T} ReLU(\textbf{W}_{\mathcal{P}_l}^{uv} [\textbf{p}_u, \textbf{q}_j ] + \textbf{b}_{\mathcal{P}_l}^{uv}),
\label{equ:equ4}
\end{equation}
\begin{equation}
   \alpha_{j}^{\mathcal{P}_l} = Softmax(e_{j}^{\mathcal{P}_l}) = 
   \frac{exp({e_{j}^{\mathcal{P}_l}})}{ {\textstyle \sum_{j'\in Y_{\mathcal{V}}(u)}} exp({e_{j'}^{\mathcal{P}_l}})},
\label{equ:equ5}
\end{equation}
where $e_{j}^{\mathcal{P}_l}$ in Equation (\ref{equ:equ4}) denotes the preference coefficient of user $u$ for item $j$ on the meta-path $\mathcal{P}_l$, the $\textbf{W}_{\mathcal{P}_l}^{uv} \in \mathbb{R}^{F \times F}$ is the weight parameters in the item perspective during the calculation process of the attention model, $\textbf{b}_{\mathcal{P}_l}^{uv} \in \mathbb{R}^{F}$ is the bias parameter, and $(\textbf{h}_{\mathcal{P}_l}^{v})^{T}$ is the learnable matrix parameter of the item perspective on the meta-path $\mathcal{P}_l$.
The \emph{Softmax} function in Equations (\ref{equ:equ5}) normalizes the preference coefficients to facilitate the fusion of the preferences of different users in groups.

According to Example \ref{example:1}, HINs contain multiple types of meta-paths.
Therefore, we calculate the users' explicit preferences on different types of meta-paths as their explicit preferences.
We accumulate the explicit preference representations of users on different types of meta-paths to obtain $\textbf{p}_u^{\mathcal{P}} \in \mathbb{R}^{F}$, as follows:
\begin{equation}
   \textbf{p}_{u}^{\mathcal{P}} = \sum_{l=1}^{|\mathcal{P}|} \textbf{p}_{u}^{\mathcal{P}_l},
\label{equ:equ6}
\end{equation}
where $|\mathcal{P}|$ denotes the number of types of meta-paths.

\subsubsection{Modeling Users' Implicit Preferences}
In HINs, we can establish the association between users using items and their dependency relationship on dependency meta-paths.
At the same time, users can establish the multi-hop interaction with items along the dependency relationship on dependency meta-paths to form implicit preference representations of users, which is the key to achieving our recommendation tasks.
For example, user $u_1$ can be associated with user $u_5$ on the dependency meta-path instance $u_1 \stackrel{1}{\rightarrow} v_{1} \mapsto v_{3} \stackrel{1}{\leftarrow} u_5$, we can model the implicit preference of user $u_1$ for item $v_3$ via the dependency relationship between $v_{1}$ and $v_{3}$.
Meanwhile, user $u_1$ can be associated with user $u_4$ on the dependency meta-path instance $u_1 \stackrel{1}{\rightarrow} v_{1} \mapsto v_{2} \stackrel{1}{\leftarrow} u_4$, we can also model the implicit preference of user $u_1$ for item $v_2$.
Therefore, we derive the implicit preference representations of users from the view of dependency-meta-paths-based via the multi-hop interaction between users and items.

In HINs, we use \textsc{PAAE} to model the multi-hop interaction between users and items on all dependency meta-paths, thereby representing the implicit preference representation of user $u$, denoted as $\textbf{p}_u^{\mathcal{PP}}$.
For the preference representation of user $u$ on a type of the dependency meta-path $\mathcal{PP}_l$, we denote it by $\textbf{p}_u^{\mathcal{PP}_l} \in \mathbb{R}^{F}$ and calculate as follows:
\begin{equation} \small
   \textbf{p}_u^{\mathcal{PP}_l} = \textsc{PAAE}(\textbf{Y}^{\mathcal{UV}}, \textbf{Y}^{\mathcal{VV}}, \mathcal{PP}_l, u) = \sum_{i \in Y_{\mathcal{V}}(u), j\in Y_{\mathcal{V}}(v)} \beta_{i,j}^{\mathcal{PP}_l} \textbf{q}_{j},
\label{equ:equ7}
\end{equation}
where the $Y_{\mathcal{V}}(u)$ is the set of items that the user $u$ interacts with item $i$ ($v_i$ is abbreviated as $i$) on the dependency meta-path $\mathcal{PP}_l$, the $Y_{\mathcal{V}}(v)$ denotes the set of the item $j$ ($v_j$ is abbreviated as $j$) that user $u$ subsequently focuses on via the dependency relationship between the items $i$ and $j$, and the $\textbf{q}_{j}$ is the embedding of item $j$.
The $\beta_{i,j}^{\mathcal{P}_l}$ is the user's attention weight to the item $j$ from the perspective of the dependency meta-path $\mathcal{PP}_l$, which indicates the latent importance of different items to user $u$.
In reality, if user $u_1$ interacts with item $v_1$ via the dependency relationship on multiple dependency meta-paths and user $u_2$ interacts with item $v_1$ on one dependency meta-path, then user $u_1$ has a higher influence than user $u_2$ in groups.
Therefore, we input the embedding $\textbf{p}_u$ of the user $u$ and the embedding $\textbf{q}_{j}$ of the item $j$ on the dependency meta-path $\mathcal{PP}_l$ into the attention model to calculate the attention weight $\beta_{i,j}^{\mathcal{PP}_l}$, as follows:
\begin{equation}
   e_{i,j}^{\mathcal{PP}_l} = (\textbf{h}_{\mathcal{PP}_l}^{v})^{T} ReLU(\textbf{W}_{\mathcal{PP}_l}^{uv} [\textbf{p}_u, \textbf{q}_j ] + \textbf{b}_{\mathcal{PP}_l}^{uv}),
\label{equ:equ8}
\end{equation}
\begin{equation} \small
   \beta_{i,j}^{\mathcal{PP}_l} = Softmax(e_{i,j}^{\mathcal{PP}_l}) = 
   \frac{exp({e_{i,j}^{\mathcal{P}_l}})}{{\textstyle \sum_{i\in Y_{\mathcal{V}}(u), j'\in Y_{\mathcal{V}}(v)}} exp({e_{i,j'}^{\mathcal{P}_l}})},
\label{equ:equ9}
\end{equation}
where $e_{i,j}^{\mathcal{PP}_l}$ in Equation (\ref{equ:equ8}) denotes the preference coefficient of user $u$ for item $j$ on the dependency meta-path $\mathcal{PP}_l$, the $\textbf{W}_{\mathcal{PP}_l}^{uv} \in \mathbb{R}^{F \times F}$ is the matrix parameters in the item perspective during the calculation of the attention model, $\textbf{b}_{\mathcal{PP}_l}^{uv} \in \mathbb{R}^{F}$ is the bias parameter, and $(\textbf{h}_{\mathcal{PP}_l}^{v})^{T}$ is the matrix parameter of the item perspective learnable on the dependency meta-path $\mathcal{PP}_l$.
The \emph{Softmax} function in Equations (\ref{equ:equ9}) normalizes the preference coefficients to facilitate the fusion of the implicit preferences of different users in groups.

According to Example \ref{example:2}, HINs contain multiple dependency meta-path types.
Therefore, we calculate the users' implicit preferences on different types of dependency meta-paths as their implicit preferences.
We accumulate the preference representations of users on different types of dependency meta-paths to obtain $\textbf{p}_u^{\mathcal{PP}} \in \mathbb{R}^{F}$, as follows:
\begin{equation}
   \textbf{p}_u^{\mathcal{PP}} = \sum_{l=1}^{|\mathcal{PP}|} \textbf{p}_u^{\mathcal{PP}_l},
\label{equ:equ10}
\end{equation}
where $|\mathcal{PP}|$ denotes the number of types of dependency meta-paths.

\subsection{User-level Recommendation}
In this section, we model users' comprehensive preferences in HINs and optimize the embedding representations between users and items to achieve the user-level recommendation task.
Therefore, we concatenate the user's preference representations on meta-paths and dependency meta-paths with their inherent features, thereby representing the comprehensive preferences of users in HINs.
Then, we perform a Multi-Layer Perceptron (\emph{MLP}) to nonlinearly model users' preference representations after concatenation to express the importance of different types of paths and the interaction between users and items.
The comprehensive explicit and implicit preference representations of the user $u$ on meta-paths and dependency meta-paths denote $\hat{\textbf{p}}_u^{\mathcal{P}} \in \mathbb{R}^{F}$ and $\hat{\textbf{p}}_u^{\mathcal{PP}} \in \mathbb{R}^{F}$, respectively, they calculate as follows:
\begin{equation}
   \hat{\textbf{p}}_u^{\mathcal{P}} = \emph{MLP}([\textbf{p}_u, \textbf{p}_u^{\mathcal{P}}]),
\label{equ:equ11}
\end{equation}
\begin{equation}
   \hat{\textbf{p}}_u^{\mathcal{PP}} = \emph{MLP}([\textbf{p}_u, \textbf{p}_u^{\mathcal{PP}}]),
\label{equ:equ12}
\end{equation}
where the [,] denotes the concatenate operation.

Then, we design a gated fusion mechanism that fuses $\hat{\textbf{p}}_u^{\mathcal{P}}$ and $\hat{\textbf{p}}_u^{\mathcal{PP}}$ to reflect the comprehensive preferences of users in the actual process. The calculation is as follows:
\begin{equation}
   \hat{\textbf{p}}_u = \eta  \odot \hat{\textbf{p}}_u^{\mathcal{P}} + (1 - \eta) \odot \hat{\textbf{p}}_u^{\mathcal{PP}},
\label{equ:equ13}
\end{equation}
\begin{equation}
   \eta = \sigma (\textbf{W}_{fusion} (\hat{\textbf{p}}_u^{\mathcal{P}} + \hat{\textbf{p}}_u^{\mathcal{PP}}) + \textbf{b}_{fusion}),
\label{equ:equ14}
\end{equation}
where $\hat{\textbf{p}}_u \in \mathbb{R}^{F}$ of Equation (\ref{equ:equ13}) reflects the overall preference representations of the user $u$ in practice, $\eta$ is the fusion ratio of the gated fusion mechanism, $\textbf{W}_{fusion} \in \mathbb{R}^{F \times F}$ is the matrix parameter, $\textbf{b}_{fusion} \in \mathbb{R}^{F}$ is the bias parameter, and $\sigma$ is the \emph{Sigmoid} activation function.

Next, we define the loss function $\mathcal{L}_u$ of user-level in group recommendations through the interaction between users and items.
Following \cite{SankarWWZYS20}, we transformed $\hat{\textbf{p}}_u$ through a fully connected network (\emph{FCN}) and normalized it by the \emph{softmax} function to produce a probability vector $\pi (\hat{\textbf{p}}_u)$ on the item set $\mathcal{V}$.
In reality, the historical interaction between users and items contains two types, the interaction $\textbf{Y}^{\mathcal{UV}}$ and multi-hop interaction $\textbf{Y}^{\mathcal{UVV}}$, which we denote them uniformly as $\bar{\textbf{Y}}^{\mathcal{UV}}$.
Then, we calculate the Kullback-Leibler (KL) divergence between the historical user-item interaction $\bar{\textbf{Y}}^{\mathcal{UV}}$
and the prediction probability $\pi (\hat{\textbf{p}}_u)$ to obtain the user's loss function $\mathcal{L}_u$, given by:
\begin{equation}
   \mathcal{L}_u = -\sum_{u \in \mathcal{U}}\frac{1}{|\bar{\textbf{y}}_{u}|} \sum_{v \in \mathcal{V}} \bar{\textbf{y}}_{uv} \; log \; \pi_{v}(\hat{\textbf{p}}_u),
\label{equ:equ15}
\end{equation}
\begin{equation}
   \pi (\hat{\textbf{p}}_u) = softmax(FCN(\textbf{W}_{v}^{u} \hat{\textbf{p}}_u)),
\label{equ:equ16}
\end{equation}
where the $\textbf{W}_{v}^{u} \in \mathbb{R}^{F \times F}$ is a learnable matrix parameter,
the $|\bar{\textbf{y}}_{u}|$ denotes the number of items that the user $u$ interacts with,
the $\bar{\textbf{y}}_{uv}$ is the specific interaction item $v$ of user $u$.

\subsection{Group-level Recommendation}
\label{sub:group}
In this section, we need to aggregate the users' preferences in each group as groups' preferences to recommend appropriate items to each group.
We consider two preference aggregators: \textsc{Meanpool} and \textsc{Attention}, to aggregate the users' preferences in groups, where the \textsc{Meanpool} mirrors the heuristics of averaging \cite{BaltrunasMR10, BerkovskyF10} and the \textsc{Attention} learn varying members' preferences in groups \cite{Cao0MAYH18, TranPTLCL19}.
We define the attention preference aggregator as follows.


\noindent\textbf{\textsc{Attention}.}
In a group, to identify the influence of different user preferences, we use an attention mechanism to calculate the weighted sum of users' preferences.
The weight parameters are learned through the attention mechanism and parameterized by \emph{MLP}.
We compute the preference representation of a group $g$ by \textsc{Attention} and denote it as $\textbf{r}_g \in \mathbb{R}^{F}$, given by:
\begin{equation}
   \textbf{r}_g = \sum_{u \in g} \gamma_{u} \; \hat{\textbf{p}}_u,
\label{equ:equ18}
\end{equation}
\begin{equation}
   o_u = \textbf{h}^T \emph{MLP}(\textbf{W}_{agg} \; \hat{\textbf{p}}_u + \textbf{b}),
\label{equ:equ19}
\end{equation}
\begin{equation}
   \gamma_{u} = Softmax(o_u) = 
   \frac{exp(o_u)}{{\textstyle \sum_{u' \in g}} exp((o_{u'})},
\label{equ:equ20}
\end{equation}
where the $\textbf{W}_{agg} \in \mathbb{R}^{F \times F}$ a matrix parameter, the $\textbf{b} \in \mathbb{R}^{F}$ is a bias parameter, $\textbf{h}^T \in \mathbb{R}^{F \times F}$ is a learnable matrix parameter between users and items, and $\gamma_{u}$ represents the influence weight of user $u$ on items that the group interacts with.

Similar to the loss function $\mathcal{L}_u$ of the user-level recommendation, we define the loss function $\mathcal{L}_g$ of the group-level recommendation during group recommendations.
We know the historical interaction between users and items contains the interaction $\textbf{Y}^{\mathcal{UV}}$ and multi-hop interaction $\textbf{Y}^{\mathcal{UVV}}$.
Therefore, we can obtain the historical interaction between groups and items, the interaction $\textbf{Y}^{\mathcal{GV}}$ and multi-hop interaction $\textbf{Y}^{\mathcal{GVV}}$, which are denoted uniformly as $\bar{\textbf{Y}}^{\mathcal{GV}}$.
We transformed $\textbf{r}_g$ through a fully connected network (\emph{FCN}) and normalized it by the \emph{softmax} function to produce a probability vector $\pi (\textbf{r}_g)$ on the item set $\mathcal{V}$.
Then, we calculate the Kullback-Leibler (KL) divergence between the historical group-item interaction $\bar{\textbf{Y}}^{\mathcal{GV}}$ and the prediction probability $\pi (\textbf{r}_g)$ to obtain the group's loss function $\mathcal{L}_g$, given by:

\begin{equation}
   \mathcal{L}_g = -\sum_{g \in \mathcal{G}}\frac{1}{|\bar{\textbf{y}}_{g}|} \sum_{v \in \mathcal{V}} \bar{\textbf{y}}_{gv} \; log \; \pi_{v}(\textbf{r}_g),
\label{equ:equ21}
\end{equation}
\begin{equation}
   \pi (\textbf{r}_g) = softmax(FCN(\textbf{W}_{v}^{g} \textbf{r}_g)),
\label{equ:equ22}
\end{equation}
where the $\textbf{W}_{v}^{g} \in \mathbb{R}^{F \times F}$ is a learnable matrix parameter,
the $|\bar{\textbf{y}}_{g}|$ denotes the number of items that the group $g$ interacts with,
the $\bar{\textbf{y}}_{gv}$ is the specific interaction item $v$ of group $g$.

\subsection{\textsc{DREAGR} Model Optimization}
\label{sub:opti}
The overall objective of the \textsc{DREAGR} model consists of two parts: the loss function $\mathcal{L}_u$ of user-level recommendation and the loss function $\mathcal{L}_g$ of group--level recommendation.
The objective of the \textsc{DREAGR} model is given by:
\begin{equation}
   \mathcal{L}(\Theta) = \mathcal{L}_u(\Theta_u) + \mathcal{L}_g(\Theta_g)
\label{equ:equ23},
\end{equation}
where $\Theta$ denotes all parameters of the \textsc{DREAGR} model, and $\Theta_u$ and $\Theta_g$ are the parameters of user-level and group-level recommendations, respectively.

We train the objective $\mathcal{L}$ of the \textsc{DREAGR} model to obtain the optimal result of the recommendation task for occasional groups.
The training process of \textsc{DREAGR} consists of two parts, and we employ the Adam \cite{KingmaB14} optimizer for training.
Firstly, the loss function $\mathcal{L}_u$ of the interaction between users and items is trained to obtain the best results for the preference representation $\hat{\textbf{p}}_u$ of user $u$ and the corresponding weight parameters $\Theta_u$.
Then, we aggregate the preference representations $\hat{\textbf{p}}_u$ of user $u$ in a group as the group's preference representations and input the weight parameters $\Theta_u$ obtained during training $\mathcal{L}_u$ into $\mathcal{L}_g$ to get the group's preference representations $\textbf{r}_g$ and the corresponding weight parameters $\Theta_g$.
Finally, we obtain the prediction probability through a predictor to denote the recommendation results.

\section{Experiments}
\label{Exper}
In this section, we conduct extensive experiments on two public datasets and present the experimental results and analysis of our \textsc{DREAGR} model.

\label{sec:exper}
\subsection{Experimental Settings}
\noindent \textbf{Datasets.}
We conduct experiments with two real-world datasets (\textbf{MOOCCube} \cite{YuLXZWFLWHLLT20} and \textbf{Movielens} \cite{Movielen}) to verify the effectiveness of our \textsc{DREAGR} model. We present the description of two datasets below:

\begin{table}
\caption{Summary statistics of three real-world datasets}
\vspace{-1.0em}
\label{table:symbol}
\begin{center}
  \begin{tabular}{lcc}
  \hline
  \textbf{Dataset}&\textbf{MOOCCube} &\textbf{Movielens} \\
  \hline
    \# Users & 17908 & 895 \\
    \# Items & 394 & 1679\\
    \# Groups & 2447 & 150\\
    \# $\mathcal{V}$-$\mathcal{V}$ dependencies & 937 & 6173 \\
    \# $\mathcal{U}$-$\mathcal{V}$ interactions & 616835 & 96464 \\
    \# $\mathcal{U}$-$\mathcal{V}$-$\mathcal{V}$ interactions & 1982499 & 16062 \\
    \# $\mathcal{G}$-$\mathcal{V}$ interactions & 93910 & 47725 \\
    \# $\mathcal{G}$-$\mathcal{V}$-$\mathcal{V}$ interactions & 100360 & 8191 \\
    Avg. \# items/user & 34.44 & 107.78 \\
    Avg. \# item-items/user & 110.70 & 17.98 \\
    Avg. \# items/group & 38.38 & 318.17 \\
    Avg. \# item-items/group & 41.01 & 54.61 \\
    Avg. group size & 7.32 & 5.97 \\
  \hline
  \end{tabular}
\end{center}
\vspace{-2.0em}
\end{table}

\noindent \textbf{Baselines.}
We evaluate the performance of \textsc{DREAGR} by comparing it with several state-of-the-art group recommendation models.
\begin{itemize}[leftmargin=*]
  \item \textbf{NCF} \cite{HeLZNHC17}. \emph{Neural Collaborative Filtering} (NCF) is a personalized recommender framework for recommending items to users. We treat a group as virtual users to utilize NCF to recommend items for a group of users.
  We adopt a predefined strategy of averaging \cite{BaltrunasMR10} to aggregate users' preferences as a group's preferences and use the Generalized Matrix Factorization (GMF) framework, the Multi-Layer Perceptron (MLP) framework, and the fusion of GMF and MLP to form the Neural Matrix Factorization (Neural Matrix Factorization (NMF) framework to achieve recommending appropriate items to a group, which are denoted as NCF-GMF, NCF-MLP, and NCF-NMF, respectively.
  \item \textbf{AGREE} \cite{Cao0MAYH18}. It addresses group representation learning with the attention network and learning the complicated interactions among groups, users, and items with NCF.
  \item \textbf{GroupIM} \cite{SankarWWZYS20}. It integrates neural preference encoders and aggregators for ephemeral group recommendation. Then it utilizes maximizing mutual information between representations of groups and group members to regularize the user-group latent space to overcome the problem of group interaction sparsity.
  \item \textbf{GBERT} \cite{ZhangZW22}. It is a pre-trained and fine-tuned method to improve group recommendations and uses BERT to enhance expression and capture learner-specific preferences. In the pre-training phase, GBERT mitigates the data sparsity problem and learns better user representations through pre-training tasks. In the fine-tuning phase, GBERT adjusts the preference representations of users and groups through influence-based moderation goals and assigns weights based on the influence of each user.
  \item \textbf{CubeRec} \cite{ChenYLNWW22}. It adaptively learns group hypercubes from user embeddings with minimal information loss in preference aggregation, measures the affinity between group hypercubes and item points via a revamped distance metric, and uses the geometric expressiveness of hypercubes to solve the issue of data sparsity.
  \item \textbf{ConsRec} \cite{WuX0J0ZY23}. It contains three novel views (including member-level aggregation, item-level tastes, and group-level inherent preferences) to provide complementary information. It integrates and balances the multi-view information via an adaptive fusion component.
\end{itemize}

\begin{table*}[h]
    \caption{Overall performance comparison on two datasets.}
    \vspace{-0.5em}
    \label{tab:result}
    \centering
    \begin{tabular}{p{1.8cm}<{\centering}ccccccccc}
    \toprule
    \multirow{3}{*}{\textbf{Datasets}} & \multirow{3}{*}{\textbf{Models}} & \multicolumn{8}{c}{\textbf{Metrics}} \\
    \cmidrule{3-10}
    &  & \multicolumn{4}{c}{\textbf{HR@N}} & \multicolumn{4}{c}{\textbf{NDCG@N}} \\
    \cmidrule{3-10}
    &  & N=5   & N=10  & N=20 & & N=5  & N=10  & N=20 &p-value\\
    \hline
    \multirow{10}{*}{\textbf{MOOCCube}}  
    &NCF-GMF &0.1704 &0.3188 &0.5840 & &0.1057 &0.1481 &0.2183& 3.42e-06  \\
    &NCF-MLP &0.2881 &0.5480 &0.7781 & &0.1817 &0.2557 &0.3054& 1.98e-04 \\
    &NCF-NMF &0.2538 &0.4716 &0.7814 &   &0.1562 &0.2284 &0.3053& 1.52e-04 \\
    &AGREE &0.3984 &0.5807 &0.7872 &   &0.2858 &0.3395 &0.3997& 1.42e-04 \\
    &GroupIM &0.8472 &0.7774 &0.6908 &   &0.8993 &0.8872 &0.8744& 2.30e-02 \\
    &BGERT &0.4253 &0.6359 &0.8067 &   &0.3569 &0.4072 &0.4859 & 1.92e-04 \\
    &CubeRec &0.6922 &0.7322 &0.8140 &   &0.6809 &0.6877 &0.7167& 2.46e-04 \\
    &ConsRec &0.8663 &0.8740 &0.9011 &   &0.8658 &0.8682 &0.8749& 1.01e-04 \\
    &\textsc{\textbf{DREAGR}} &\textbf{0.9270} &\textbf{0.9191} &\textbf{0.9062} &   &\textbf{0.9320} &\textbf{0.9206} &\textbf{0.9056} & - \\
    &Min. improvement rate &\textbf{7.01\%} &\textbf{5.16\%} &\textbf{0.57\%}  &   &\textbf{3.64\%} &\textbf{3.76\%} &\textbf{3.51\%}& -\\
    \hline
    \multirow{10}{*}{\textbf{Movielens}}  
    &NCF-GMF &0.0733 &0.1333 &0.2667 &  &0.0426 &0.0657 &0.1019 & 3.47e-09 \\
    &NCF-MLP &0.2067 &0.2600 &0.4200 &   &0.1253 &0.1388 &0.1840 & 1.80e-07 \\
    &NCF-NMF &0.1533 &0.3000 &0.4933 &   &0.0864 &0.1413 &0.1955 & 2.77e-06 \\
    &AGREE &0.5704 &0.7094 &0.8152 &   &0.4311 &0.4816 &0.5062 & 3.99e-03 \\
    &GroupIM &0.8303 &0.7939 &0.7697 &   &0.8326 &0.8123 &\textbf{0.7926} & 0.5145 \\
    &BGERT &0.6092 &0.7578 &0.8432 &   &0.5012 &0.5623 &0.6284 & 0.0116 \\
    &CubeRec &0.4241 &0.4009 &0.4359 &   &0.4465 &0.4197 &0.4212& 7.34e-010 \\
    &ConsRec &0.6695 &0.7359 &\textbf{0.8469} &   &0.6501 &0.6711 &0.6991& 0.0106 \\
    &\textsc{\textbf{DREAGR}} &\textbf{0.8545} &\textbf{0.8182} &0.7576 &   &\textbf{0.8608} &\textbf{0.8332} &0.7851& - \\
    &Min. improvement rate &\textbf{2.91\%} &\textbf{3.06\%} &\textbf{-10.54\%} &   &\textbf{3.39\%} &\textbf{2.57\%} &\textbf{-0.95\%} &- \\
    \bottomrule
    \end{tabular}
\vspace{-1.5em}
\end{table*}

\noindent \textbf{Parameter Settings.}
To evaluate the recommendation performance of \textsc{DREAGR} and comparison models, we divided our datasets into training, validation, and testing sets in a 7:1:2 ratio.
Regarding the parameter settings of baseline models, we follow the optimal parameters in the original paper of these models and fine-tune them based on our datasets to ensure the optimal performance of these models.
For the parameters of the \textsc{DREAGR} model, we set its learning rate as \{0.00005, 0.0001, 0.0005, 0.001, 0.005, 0.01, 0.05\}, the embedding dimensions of users and items as \{32, 64, 128, 256, 512, 1024\}, the range of Batch\_size is \{32, 64, 128, 256, 512, 1024\}, which is the number of training samples used at one time during the training process, and weight decay (denoted by $\lambda$) is \{0.0, 0.001, 0.005, 0.01, 0.05\}, to analyze the performance of \textsc{DREAGR} under different combinations of parameters.
We set the epoch during the training process of \textsc{DREAGR} and its variants to 50.
The learning rates during user pre-training are 0.01 and 0.005 on the \textbf{MOOCCube} and \textbf{Movielens} datasets, respectively.

\noindent \textbf{Evaluation Metrics.}
To evaluate the performance of group recommendations, we employ Hit Ratio (HR@N) and Normalized Discounted Cumulative Gain (NDCG@N) as evaluation metrics, where N is 5, 10, and 20.
The formula for these two evaluation metrics is as follows:
\begin{equation}
   HR@N = \frac{Number of Hit@N}{|\mathcal{O}_{test}|},
\label{equ:equ24}
\end{equation}
\begin{equation}
   NDCG@N = \frac{DCG@N}{IDCG@N},
\label{equ:equ25}
\end{equation}
where the \emph{Number of Hit@N} in Equation (\ref{equ:equ24}) represents the number of instances $N$ \emph{Hit} in the test dataset, and $|\mathcal{O}_{test}|$ indicates the number of instances in the test dataset.
The DCG in Equation (\ref{equ:equ25}) is the Discounted Cumulative Gain.
The IDCG represents the user's favorite item in the recall set and denotes the ideal maximum DCG value.

    

\subsection{Effectiveness}
We obtain the optimal performance of our \textsc{DREAGR} model based on the optimal parameter combination under two evaluation metrics (HR@N and NDCG@N, N=5, 10, 20) in Section \ref{sub:parameter}.
Then, we compare our \textsc{DREAGR} model with the state-of-the-art group recommendation models.
Table \ref{tab:result} shows the best performance of all models under different settings in terms of
HR@N and NDCG@N on the \textbf{MOOCCube} and \textbf{Movielens} datasets.

Compared to other models, the performance of NCF-based models is almost the worst on these two datasets, as they obtain group preferences through the mean aggregator and cannot learn the differences in the personal preferences of group members.
However, the metric HR@20 of the \textsc{NCF-NMF} model is even better than those of the GroupIM (on the \textbf{MOOCCube} dataset) and CubeRec (on the \textbf{Movielens} dataset) models.
This situation occurs because the \textsc{NCF-NMF} model obtains stable preference information of users based on the pre-training process of \textsc{NCF-GMF}.
We can see from Table \ref{tab:result} that the HR@N and NDCG@N (N=5, 10) of the \textsc{NCF-MLP} model are optimal among NCF-based models, while the HR@20 and NDCG@20 of the \textsc{NCF-NMF} model are optimal.
Strangely, the \textsc{NCF-NMF} model, which integrates the \textsc{GMF} and \textsc{MLP}, cannot have an advantage in all evaluation metrics.

When compared with the \textsc{AGREE}, GroupIM, \textsc{GBERT}, CubeRec, and ConsRec models, we can find that the HR@N and NDCG@N (N=5, 10, 20) of our \textsc{DREAGR} model outperforms them on the \textbf{MOOCCube} dataset.
We can find that the HR@N and NDCG@N (N=5, 10) of our \textsc{DREAGR} model are optimal on the \textbf{Movielens} dataset, while its HR@20 and NDCG@20 are not best.
Although the HR@20 of the \textsc{AGREE} model is superior to \textsc{DREAGR} on the \textbf{Movielens} dataset, the calculation time is very high due to \textsc{AGREE} reading only one pair of interaction information between groups and items for calculation each time.
In addition, the HR@20 of our \textsc{DREAGR} model is weaker than the \textsc{GBERT} and ConsRec models.
One reason is that the number of \emph{Implicit interactions} between groups and items (8,191 in Table \ref{tab:result}) in the \textbf{Movielens} dataset is much lower than their \emph{Explicit interactions} (47,725 in Table \ref{tab:result}), resulting in insufficient capture of corresponding implicit preferences in  \textsc{DREAGR}.
Another reason is that \textsc{GBERT} assigns weight to adjust the preference representation of users and groups according to users' influence in the fine-tuning phase, and \textsc{GBERT} learns to realize an efficient and expressive member-level aggregation via hypergraph.
Surprisingly, the HR@20 and NDCG@20 of our \textsc{DREAGR} model are weaker than GroupIM, indicating that GroupIM can overcome group interaction sparsity by group-adaptive preference prioritization.

Although the HR@20 and NDCG@20 of our \textsc{DREAGR} model are not the best on the \textbf{Movielens} dataset, other evaluation metrics are optimal on these two datasets, which indicates that our \textsc{DREAGR} model is still effective.
We calculate the minimum improvement rate when comparing the \textsc{DREAGR} model with the optimal comparison model on different evaluation metrics, which more intuitively shows the performance improvement of our \textsc{DREAGR} model.
In addition, we calculate that all the p-values between \textsc{DREAGR} and baselines are much smaller than 0.05 (except for GroupIM on the \textbf{Movielens} dataset), which indicates that the improvements are statistically significant.

\noindent\textbf{\underline{\textit{Insight}}:}
(1) We can find that the HR@N and NDCG@N of the GroupIM and \textsc{DREAGR} models decrease with the increase of N.
One reason is that there is a ``Long-tail Effect'' (Fig. \ref{fig:stat}) in the interactions between groups and items in our dataset, resulting in an uneven distribution of interactions in practice.
Another reason is that the GroupIM and \textsc{DREAGR} models do not rely on the historical interaction information of a group when recommending items to it, while other models do.
(2) We find that the performance of our \textsc{DREAGR} model on the \textbf{MOOCCube} dataset is better than that of the \textbf{Movielens} dataset.
The reason is that there is only one type of meta-path $(i.e., \mathcal{U} \stackrel{1}{\rightarrow} \mathcal{V} \stackrel{1}{\leftarrow} \mathcal{U})$ and dependency meta-path $(i.e., \mathcal{U} \stackrel{1}{\rightarrow} \mathcal{V}_{i'} \mapsto \mathcal{V}_{j'} \stackrel{1}{\leftarrow} \mathcal{U})$ in the \textbf{Movielens} dataset, 
and there are three types in the \textbf{MOOCCube} dataset (see the \emph{Example} \ref{example:1} and \ref{example:2}).
Therefore, we can not capture more potential preference information of groups in the \textbf{Movielens} dataset.
In addition, the number of \emph{Implicit interactions} on the dependency meta-path in the \textbf{Movielens} dataset is much lower than the explicit interactions on the meta-path, resulting in the inability to capture more implicit preferences for groups.

\begin{figure}[t]
\centering
  \subfloat[\textbf{MOOCCube}]{
    \centering
	\includegraphics[scale=0.166]{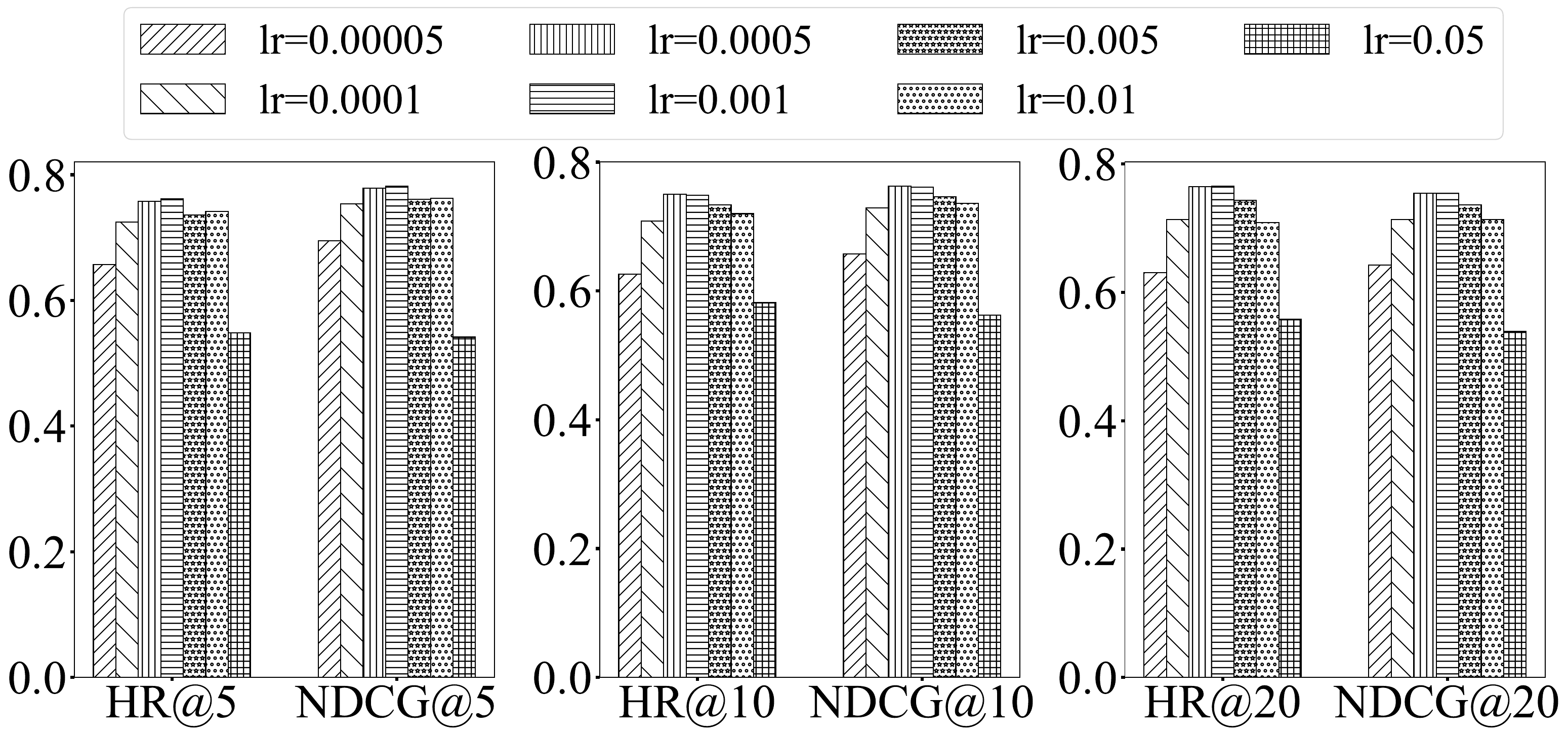}}
	\\
  \subfloat[\textbf{Movielens}]{
    \centering
	\includegraphics[scale=0.166]{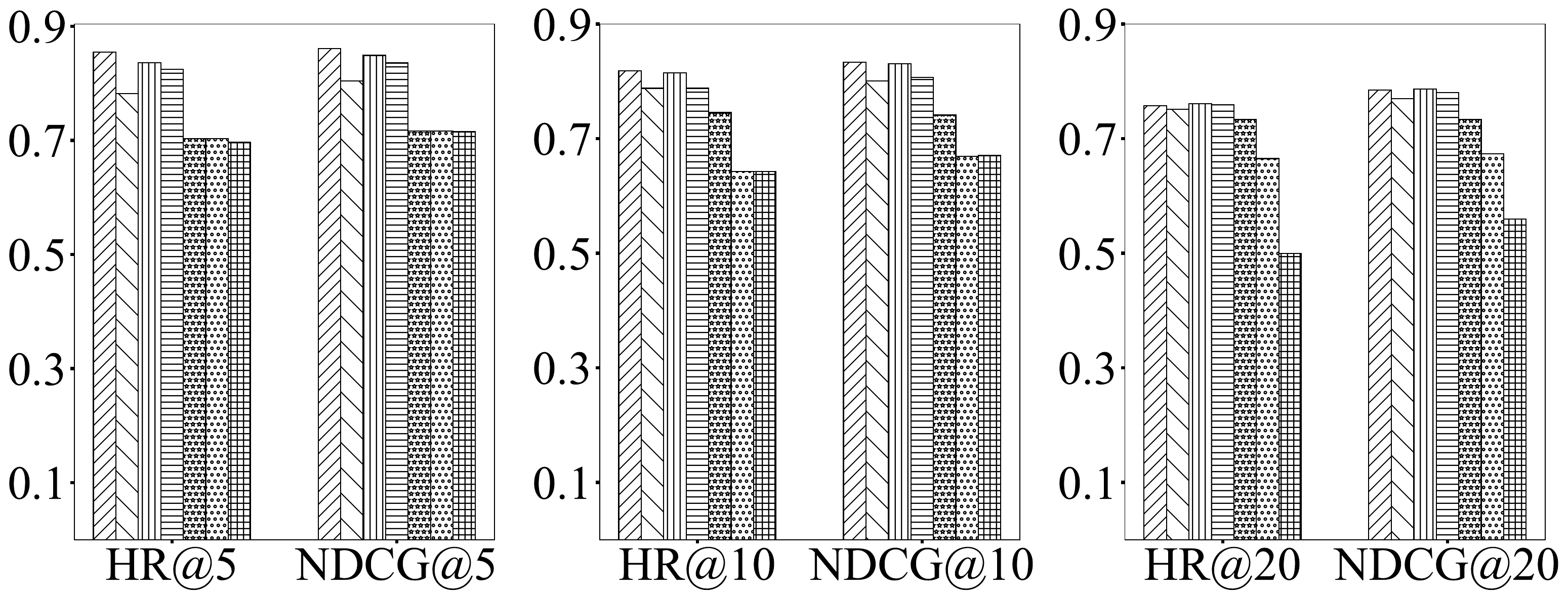}}
\vspace{-1.0em}
\caption{The performance changes of \textsc{DREAGR} as the learning rate increases.}
\vspace{-2.0em}
\label{fig:lr}
\end{figure}

\subsection{Parameter Analysis of \textsc{DREAGR}}
\label{sub:parameter}
\vspace{-0.5em}
In this section, we tune the four critical parameters of the \textsc{DREAGR} model, learning rate, the embedding dimensions of users and items, the number of training samples (i.e., Batch\_size), and the weight decay, by the evaluation metrics HR@N and NDCG@N (N=5, 10, 20) to select the optimal combination of corresponding parameters.

\subsubsection{Influence Analysis of Learning Rate}
\label{ialr}
To analyze the influence of learning rate on the \textsc{DREAGR} model, we set the embedding dimensions (users and items) to 64, the Batch\_size to 64, and the weight decay to 0.001, then set the learning rate range to \{0.00005, 0.0001, 0.0005, 0.001, 0.005, 0.01, 0.05\}.
The variation of the evaluation metrics HR@N and NDCG@N of the \textsc{DREAGR} model with increasing learning rates on the \textbf{MOOCCube} and \textbf{Movielens} datasets is shown in Fig. \ref{fig:lr}.


We can find that the evaluation metrics HR@N and NDCG@N of the \textsc{DREAGR} model show a trend of increasing and then decreasing with increasing the learning rate on the \textbf{MOOCCube} dataset, as shown in Fig. \ref{fig:lr}.
The HR@5 and NDCG@5 of the \textsc{DREAGR} model are optimal when the learning rate is 0.001.
the HR@10 and NDCG@10 of the \textsc{DREAGR} model are optimal when the learning rate is 0.0005.
The HR@20 of the \textsc{DREAGR} model is optimal when the learning rate is 0.001, while NDCG@20 of the \textsc{DREAGR} model is almost the same when the learning rate is 0.001 and 0.0005.
Therefore, we select the \textsc{DREAGR} model to have a learning rate of 0.001 on the \textbf{MOOCCube} dataset.
On the \textbf{Movielens} dataset, we find that the evaluation metrics HR@N and NDCG@N of the \textsc{DREAGR} model decreases consistently with increasing the learning rate. We select the \textsc{DREAGR} model with a learning rate of 0.00005 on the \textbf{Movielens} dataset.

\subsubsection{Influence Analysis of Embedding Dimension}
\label{iaed}
\begin{figure}
\centering
  \subfloat[\textbf{MOOCCube}]{
    \centering
	\includegraphics[scale=0.165]{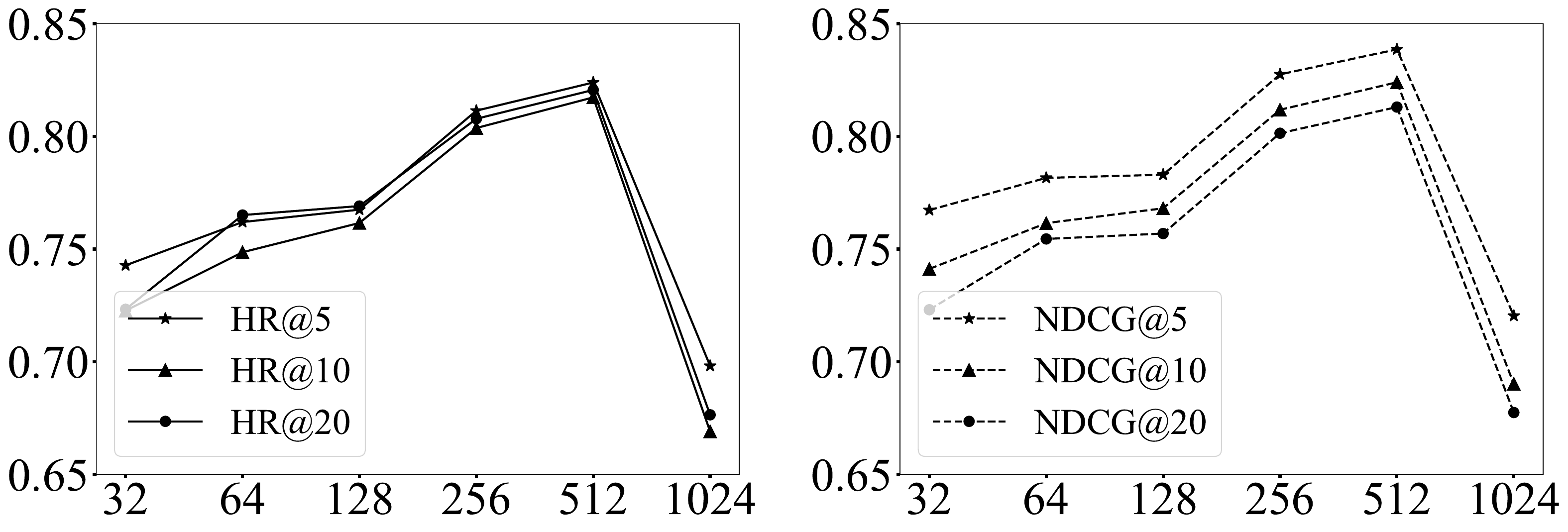}}
	\\
  \subfloat[\textbf{Movielens}]{
    \centering
	\includegraphics[scale=0.165]{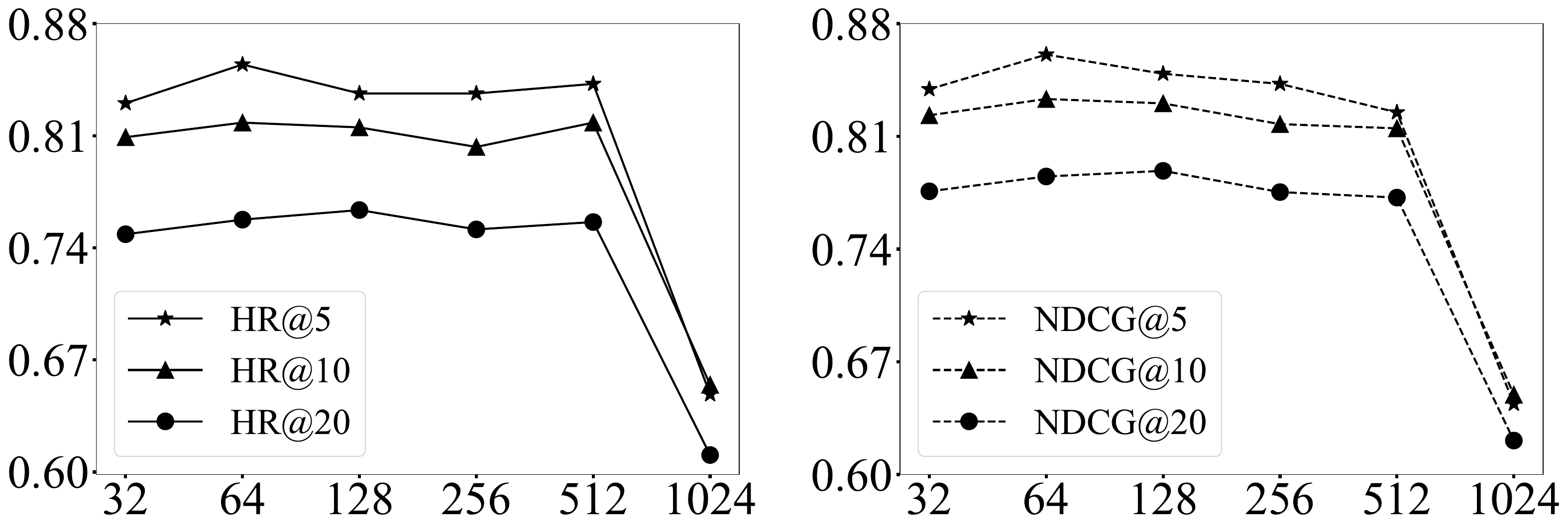}}
\vspace{-1.0em}
\caption{The performance changes of \textsc{DREAGR} as the embedding dimensions.}
\vspace{-1.5em}
\label{fig:emb}
\end{figure}

To test the influence of the embedding dimensions in the \textsc{DREAGR} model, we set the embedding dimensions of users and items during low-dimensional transformations to \{32, 64, 128, 256, 512, 1024\}.
We select the learning rate as the optimal value on both datasets (Section \ref{ialr}), set the Batch\_size as 64, and set weight decay as 0.001.
The change of the evaluation metrics HR@N and NDCG@N of the \textsc{DREAGR} model with increasing the embedding dimensions on the \textbf{MOOCCube} and \textbf{Movielens} datasets are shown in Fig. \ref{fig:emb}.


We can see that the evaluation metrics HR@N and NDCG@N of the \textsc{DREAGR} model show increasing and then decreasing with increasing the embedding representation dimensions (users and items) on the \textbf{MOOCCube} and \textbf{Movielens} datasets, as shown in Fig. \ref{fig:emb}.
The \textsc{DREAGR} model achieves optimal value for the embedding representation dimensions is 512 on the \textbf{MOOCCube} dataset, while its optimal value for the embedding representation dimensions is 64 on the \textbf{Movielens} dataset.
Therefore, we select the embedding representation dimensions of the \textsc{DREAGR} model for the corresponding low-dimensional of users and items on the \textbf{MOOCCube} and \textbf{Movielens} datasets to be 512 and 64, respectively.

\subsubsection{Influence Analysis of Batch\_size}
\label{iants}
\begin{figure}
\centering
  \subfloat[\textbf{MOOCCube}]{
    \centering
	\includegraphics[scale=0.165]{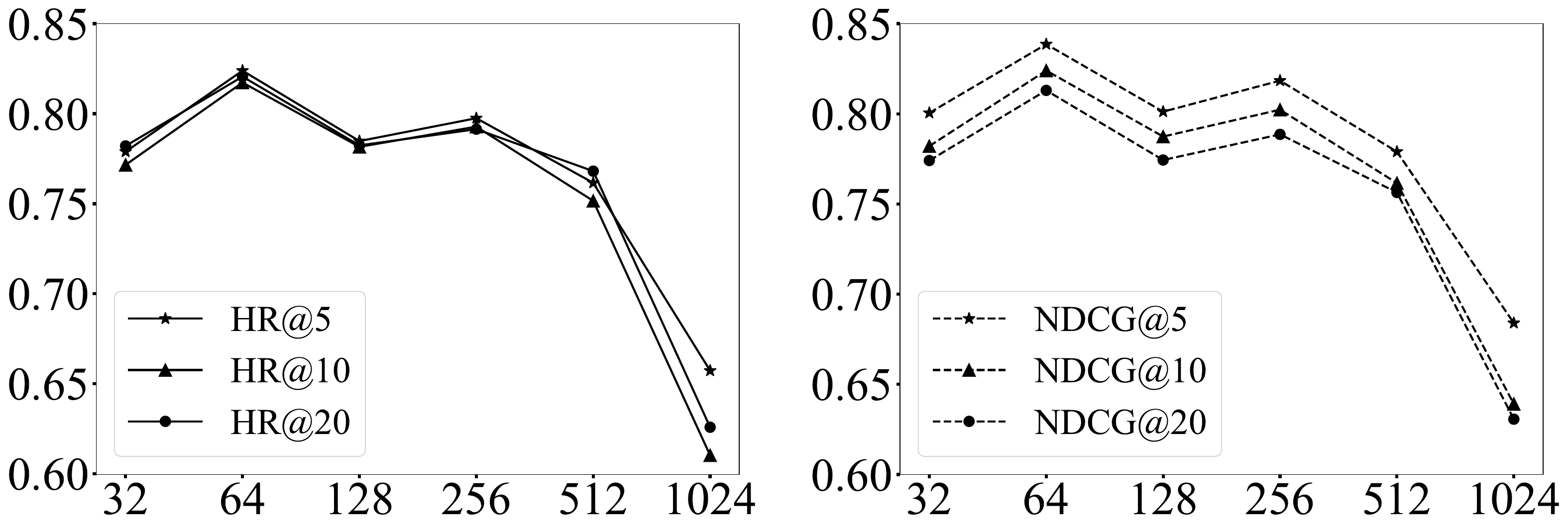}}
	\\
  \subfloat[\textbf{Movielens}]{
    \centering
	\includegraphics[scale=0.165]{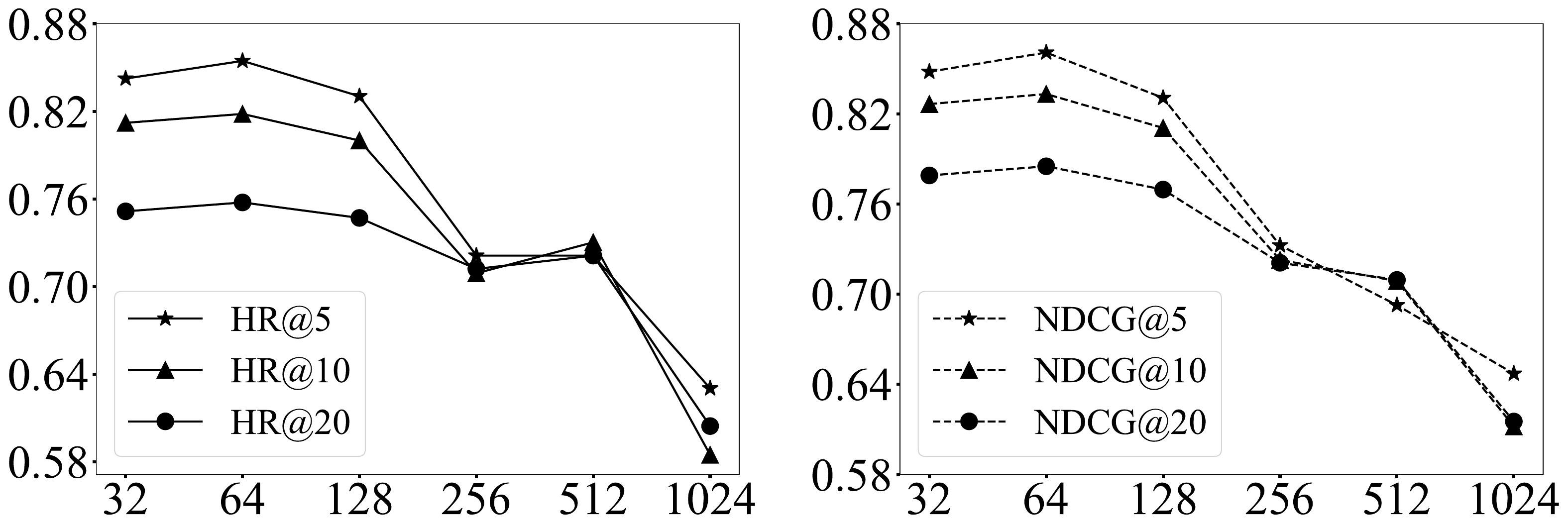}}
\vspace{-1.0em}
\caption{The performance changes of \textsc{DREAGR} as Batch\_size.}
\label{fig:batch}
\vspace{-1.5em}
\end{figure}

To test the influence of Batch\_size (the number of training samples) in the \textsc{DREAGR} model, we set its range to \{32, 64, 128, 256, 512, 1024\}.
We select the optimal value of learning rates (Section \ref{ialr}) and the embedding representation dimensions (Section \ref{iaed}) on the \textbf{MOOCCube} and \textbf{Movielens} datasets and then set the weight decay to 0.001.
The change of two evaluation metrics HR@N and NDCG@N of the \textsc{DREAGR} model with increasing Batch\_size on the \textbf{MOOCCube} and \textbf{Movielens} datasets, is shown in Fig. \ref{fig:batch}.

From Fig. \ref{fig:batch}, we can see that the evaluation metrics HR@N and NDCG@N of the \textsc{DREAGR} model show a trend of first increasing and then decreasing with the increase of the Batch\_size on the \textbf{MOOCCube} and \textbf{Movielens} dataset.
On these two datasets, the HR@N and NDCG@N of the \textsc{DREAGR} model are optimal when Batch\_size is 64.

\subsubsection{Influence Analysis of Weight Decay}
\begin{figure}[t]
\centering
  \subfloat[\textbf{MOOCCube}]{
    \centering
	\includegraphics[scale=0.166]{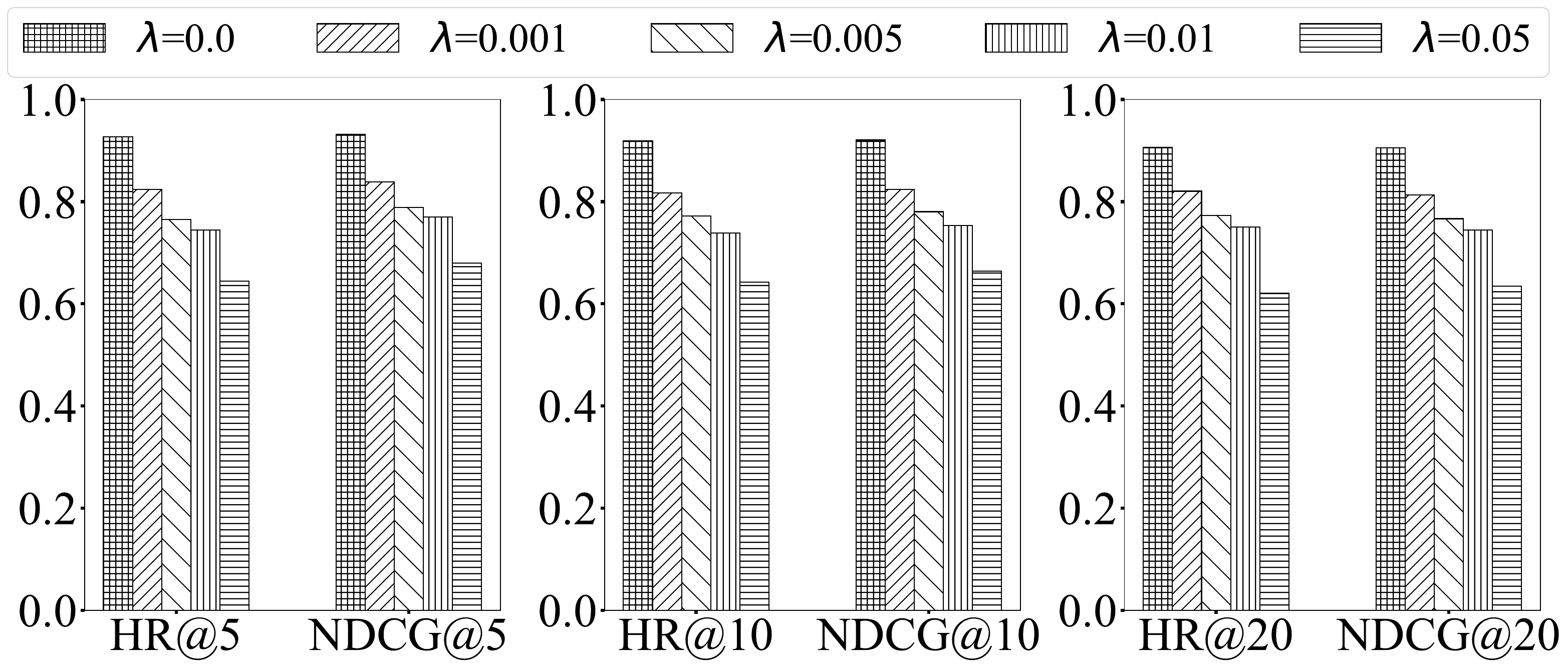}}
	\\
  \subfloat[\textbf{Movielens}]{
    \centering
	\includegraphics[scale=0.166]{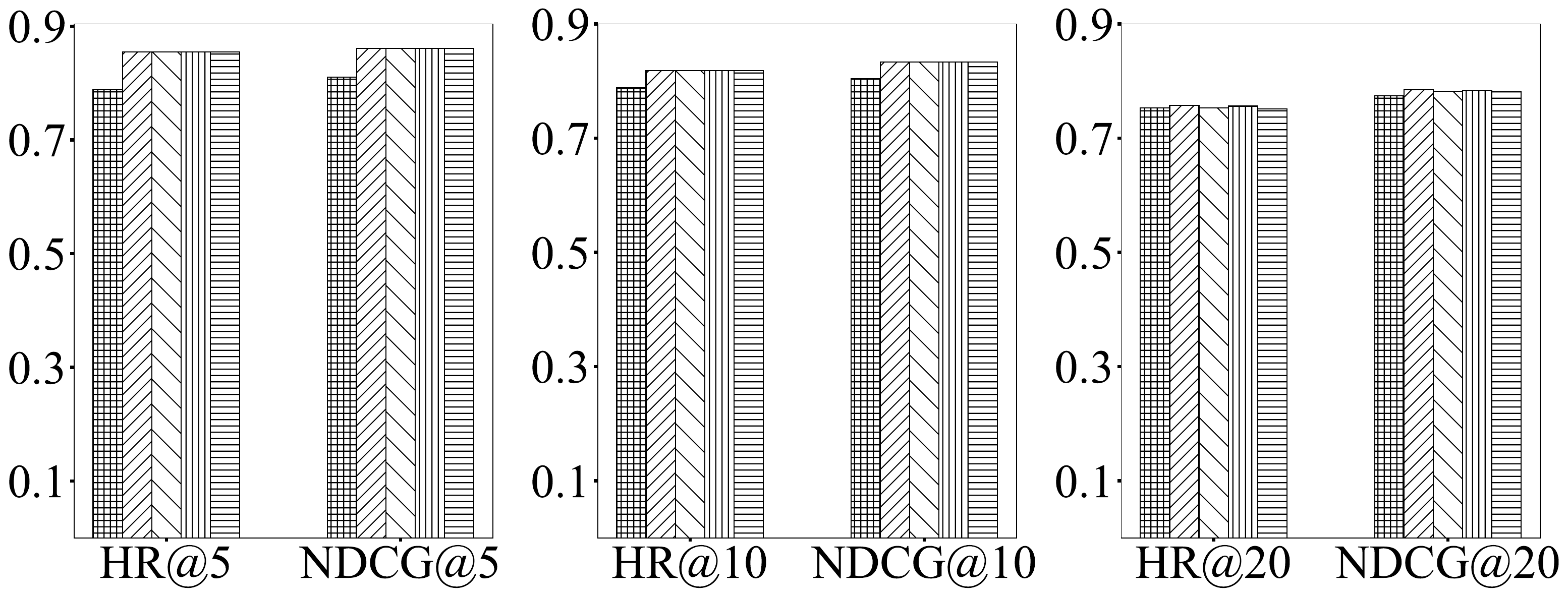}}
\vspace{-0.8em}
\caption{Performance changes of \textsc{DREAGR} as weight decay.}
\label{fig:wd}
\vspace{-1.5em}
\end{figure}

To analyze the influence of weight decay in the \textsc{DREAGR} model, we set its range to \{0.0, 0.001, 0.005, 0.01, 0.05\}.
We select the optimal value of learning rates (Section \ref{ialr}), the embedding representation dimensions (Section \ref{iaed}), and the Batch\_size (Section \ref{iants}) on the \textbf{MOOCCube} and \textbf{Movielens} datasets.
The variation of two evaluation metrics HR@N and NDCG@N of the \textsc{DREAGR} model with increasing weight decay on the \textbf{MOOCCube} and \textbf{Movielens} datasets, is shown in Fig. \ref{fig:wd}.

We can find that the evaluation metrics HR@N and NDCG@N of the \textsc{DREAGR} model on the \textbf{MOOCCube} dataset are decreasing as the value of weight decay increases, as shown in Fig. \ref{fig:wd}.
The HR@N and NDCG@N of the \textsc{DREAGR} model are optimal when the value of weight decay is 0 on the \textbf{MOOCCube} dataset.
On the \textbf{Movielens} dataset, we can see that the values of the evaluation metrics HR@N and NDCG@N of the \textsc{DREAGR} model are the same when the value of weight decay is greater than 0, especially HR@N and NDCG@N (n=5, 10).
However, we can find that the evaluation metrics HR@20 and NDCG@20 of the \textsc{DREAGR} model slightly outperform the values under other weight decay when the weight attenuation value is 0.001.
Therefore, we select the weight decay of the \textsc{DREAGR} model as 0.001 on the \textbf{Movielens} dataset.


\subsection{Ablation study}
In this section, we utilize an ablation study to verify the effectiveness of different modules of the \textsc{DREAGR} model, such as the meta-path, the dependency meta-path, the attention mechanism, and the pre-training of users.
We remove the user's pre-training, meta-path, and dependency meta-path and use a mean aggregator instead of the attention aggregator to design four variants of the \textsc{DREAGR} model, then compare them by experiments to show that they are effective.
The variants of \textsc{DREAGR} are shown below.

\begin{itemize}[leftmargin=*]
  \item \textsc{DREAGR-RPT} (Remove Pre-training of users): This variant is formed by removing the user's pre-training in the \textsc{DREAGR} model, i.e., expurgating the relevant weight matrixes and the user's optimal preferences obtained from $\mathcal{L}_u$ in Equation (\ref{equ:equ23}) during the training process.
  
  \item \textsc{DREAGR-RDMP} (Remove Dependency Meta-Paths): This variant denotes that the \textsc{DREAGR} model removes users' preference information on dependency meta-paths, i.e., we only consider users' preference information on meta-paths in the \textsc{DREAGR} model.
  
  \item \textsc{DREAGR-RMP} (Remove Meta-Paths): This variant denotes that the \textsc{DREAGR} model removes users' preference information on meta-paths, i.e., we only consider users' preference information on dependency meta-paths in the \textsc{DREAGR} model.
  
  \item \textsc{DREAGR-RAA} (Replace Attention Aggregator):
  This variant indicates that we use a mean aggregator instead of an attention aggregator in the \textsc{DREAGR} model. We mean users' preferences in a group as the group's preferences without considering the different influences of users.
\end{itemize}

\begin{figure}[t]
\centering
  \subfloat[\textbf{MOOCCube}]{
    \centering
	\includegraphics[scale=0.149]{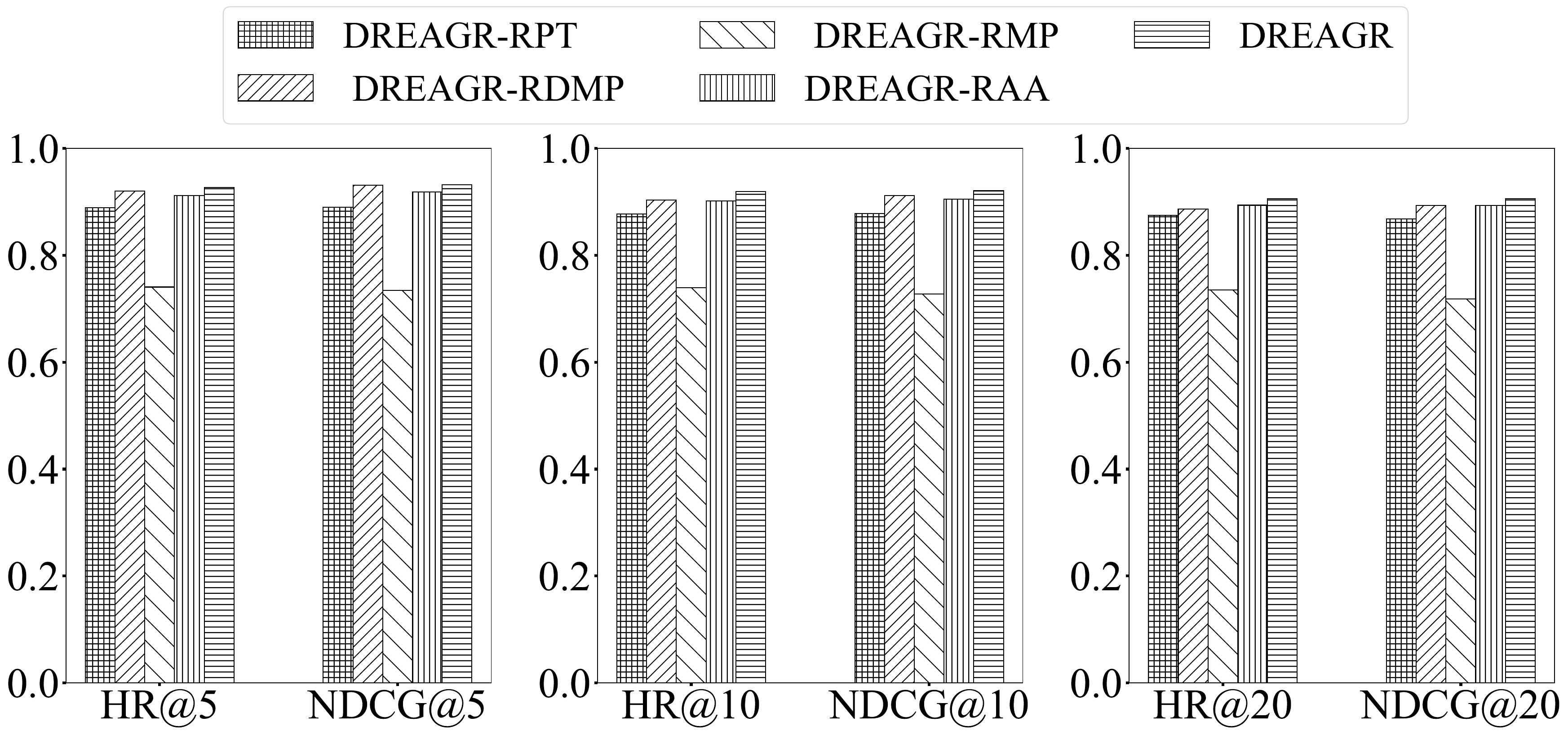}}
	\\
  \subfloat[\textbf{Movielens}]{
    \centering
	\includegraphics[scale=0.149]{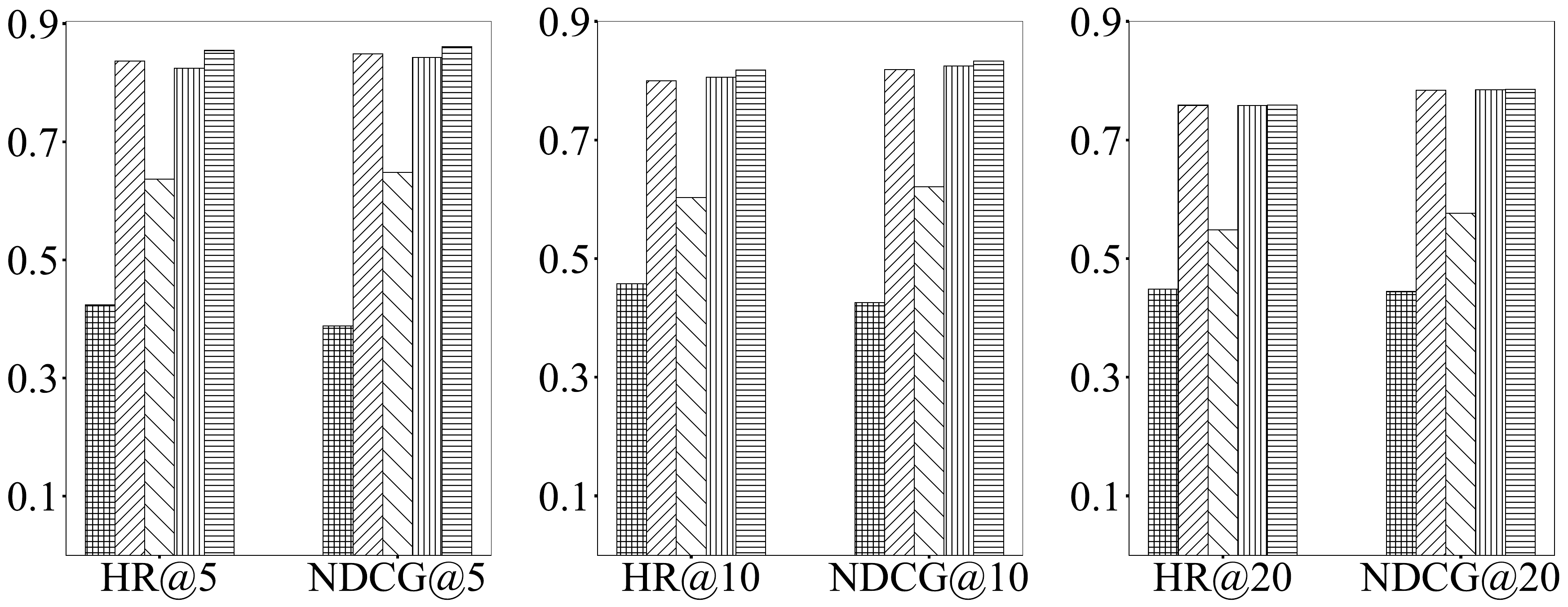}}
\vspace{-0.8em}
\caption{Results of ablation study with \textsc{DREAGR}.}
\label{fig:alb}
\vspace{-2.0em}
\end{figure}

We obtain the experimental results of the \textsc{DREAGR} model and its variants on the evaluation metrics HR@N and NDCG@N (N =5, 10, 20), as shown in Fig. \ref{fig:alb}.
Compared to the \textsc{DREAGR-RPT} model on two datasets, we know that users' pre-training can effectively improve the overall performance of our \textsc{DREAGR} model, which indicates that obtaining weight information of users during the pre-training process is effective.
Compared to the \textsc{DREAGR-RDMP} and \textsc{DREAGR-RMP} models, we find that fusing users' preferences on meta-paths and dependency meta-paths can improve the overall performance of our \textsc{DREAGR} model.
We can see that the interaction between users and items on meta-paths has a higher impact on our \textsc{DREAGR} model than on dependency meta-paths when comparing \textsc{DREAGR-RDMP} and \textsc{DREAGR-RMP}.
The reason is that the richer meta-path information in groups can help \textsc{DREAGR-RPMP} capture more users' preference information.
In practice, the interaction of our constructed users with items on dependency meta-paths is also effective and can recommend items needed for groups in the next stage.
Compared to the \textsc{DREAGR-RAA} model, since the attention aggregator in the \textsc{DREAGR} model can capture different preference information among users and aggregate it into the group's preferences, the overall performance of the \textsc{DREAGR} model is higher than the mean aggregator in the \textsc{DREAGR-RAA} model.

As shown in Fig. \ref{fig:alb}, we can see that the \textsc{DREAGR} model outperforms all its variants on the \textbf{MOOCCube} dataset for the evaluation metrics HR@N and NDCG@N (N =5, 10, 20).
On the \textbf{Movielens} dataset, we find that the \textsc{DREAGR} model outperforms all its variants for HR@N and NDCG@N (N =5, 10).
Strangely, the \textsc{DREAGR} model has almost no difference in the metrics HR@20 and NDCG@20 compared to its variants \textsc{DREAGR-RDMP} and \textsc{DREAGR-RAA}.
Since the interaction between groups and items in the \textbf{Movielens} dataset is uneven and sparse, we speculate that these three models centralized recommending certain items to groups in the recommendation process as the increase of N.
Nevertheless, this ablation study proves that combining different module contents in \textsc{DREAGR} is available on these two datasets and shows the effectiveness of module contents of our \textsc{DREAGR} model in group recommendations.

\subsection{Case study}
\begin{figure}[t]
\centering
  \includegraphics[width=\linewidth]{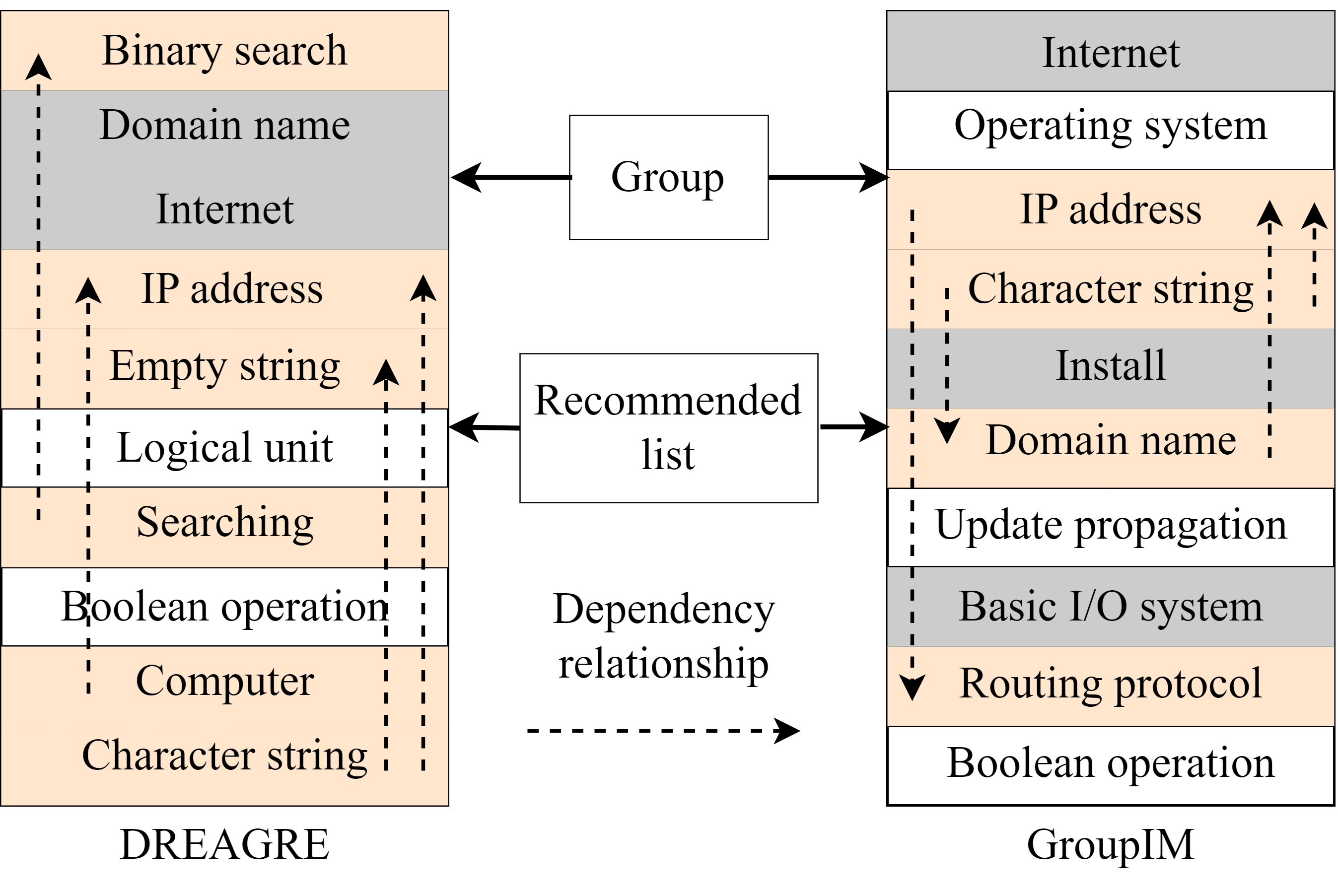}
\vspace{-2.0em}
\caption{Comparison of case studies on recommendation results between the \textsc{DREAGR} and GroupIM models.}
\label{fig:case}
\vspace{-1.5em}
\end{figure}

To further demonstrate the effectiveness of our proposed \textsc{DREAGR} model, we conduct a case study on the \textbf{MOOCCube} dataset in this section.
Since the loss function of the recommendation process for \textsc{DREAGR} and GroupIM is the same, we compare their recommendation results.
We randomly select the group and obtain the top-10 recommended list using the evaluation metric HR@N according to \textsc{DREAGR} and GroupIM, as shown in Fig. \ref{fig:case}.
The gray box indicates items that failed the recommendation, and the black dashed arrow indicates the dependency relationship between concepts within the orange box in the recommendation list.
We can intuitively observe that the \textsc{DREAGR} and GroupIM models generate different results.
We can see that \textsc{DREAGR} recommends more related items than GroupIM because we fuse users' explicit and implicit preferences.
From the recommendation results of the group, we know that the knowledge concepts currently learned by users of the group are computer fundamentals in MOOCs.
We provide the dependency relationships between items in the recommendation results, indicating that we can mine users' implicit preferences based on the user-item interactions on dependency meta-paths.
Therefore, we can briefly explain why certain items are recommended to the group based on the dependency relationships in practice.

\section{Conclusions and Future Work}
\label{conc}
In this paper, we investigated and addressed the problem of interaction sparsity and preference aggregation in the recommendation task of occasional groups.
We proposed a Dependency Relationships-Enhanced Attention Group Recommendation (\textsc{DREAGR}) model to recommend suitable items to a group of users by alleviating interaction sparsity and aggregating user preferences.
Specifically, to alleviate the problem of sparse interaction in occasional groups, we introduce the dependency relationships between items as side information to enhance the user/group-item interaction.
To model users' preferences, we defined meta paths and dependent meta paths in HINs and proposed a Path-Aware Attention Embedding (PAAE) method to learn users' preferences when interacting with items on different types of paths.
We conducted experiments on two datasets to evaluate the performance of \textsc{DREAGR}, and the experimental results validated the superiority of \textsc{DREAGR} by comparing it with state-of-the-art group recommendation models.

We have addressed the problems of interaction sparsity and preference aggregation in the recommendation task of occasional groups, and we will further investigate a significant problem in group recommendations: the fairness problem \cite{LinZZGLM17}.
Unfair attention brought a rich-get-richer problem and became a barrier for unpopular services to startups \cite{3604558}.
Recently, the fairness problem in recommendations has been investigated in many domains, such as Job recommendations \cite{LambrechtT19} and Book recommendations \cite{EkstrandTKMK18}.
\vspace{-1.0em}





\ifCLASSOPTIONcaptionsoff
  \newpage
\fi

\balance
\bibliographystyle{IEEEtranN}
\bibliography{ref}

\end{document}